\documentclass[pra, aps, twocolumn,amssymb,amsmath,amsbsy, groupedaddress, superscriptaddress]{revtex4-1}
\usepackage{slashed}
\usepackage[pdftex]{graphicx}
\usepackage{graphicx}
\usepackage{epstopdf}
\usepackage{float}
\usepackage{subcaption}
\usepackage{enumerate}
\usepackage{chngcntr}
\usepackage[usenames, dvipsnames]{color}
\usepackage{bm}
\usepackage{amsfonts}
\usepackage{dsfont}
\usepackage{lipsum}
\usepackage[utf8]{inputenc}
\usepackage[T1]{fontenc}
\usepackage{bbold}
\usepackage{times}
\usepackage{color}
\usepackage{pifont}
\usepackage{tikz}
\usepackage{colortbl}
\usepackage{fancybox}
\usepackage{epstopdf}
\usepackage{latexsym}
\usepackage{amssymb}
\usepackage{hyperref}
\hypersetup{backref,
colorlinks=true,
urlcolor=blue,
linkcolor=blue,
linktoc=page,
citecolor=blue}

\everymath{\displaystyle} 

\begin{document}

\title{Ultimate precision of joint parameter estimation under noisy Gaussian environment}

\author{Lahcen Bakmou}\email{baqmou@gmail.com} \email{lahcen_bakmou@um5.ac.ma}
\affiliation{LPHE-Modeling and Simulation, Faculty of Sciences, Mohammed V University, Rabat, Morocco.}
\author{Mohammed Daoud} \email{m_daoud@hotmail.com}
\affiliation{Department of Physics , Faculty of Sciences, University Ibn Tofail, Kenitra, Morocco.}
%\pacs{............}

\begin{abstract}
	
The major problem of multiparameter quantum estimation theory is to find an ultimate measurement scheme to go beyond the standard quantum limits that each quasi-classical estimation measurement is limited by. Although, in some specifics quantum protocols without environmental noise, the ultimate sensitivity of a multiparameter quantum estimation can beat the standard quantum limit. However, the presence of noise imposes limitations on the enhancement of precision due to the inevitable existence of environmental fluctuations. Here, we address the motivation behind the usage of Gaussian quantum resources and their advantages in reaching the standard quantum limits under realistic noise. In this context, our work aims to explore the ultimate limits of precision for the simultaneous estimation of a pair of parameters that characterize the displacement channel acting on Gaussian probes and subjected to open dynamics. More precisely, we focus on a general two-mode mixed squeezed displaced thermal state, after reducing it to various Gaussian probes states, like; a two-mode pure squeezed vacuum, two-mode pure displaced vacuum, two-mode mixed displaced thermal, two-mode mixed squeezed thermal. To study the ultimate estimation precision, we evaluate the upper and bottom bound of HCRB in various cases. We find that when the entangled states, two-mode pure squeezed vacuum and two-mode mixed squeezed thermal, are employed as probes states, the upper and bottom bound of HCRB beats the standard quantum limit in the presence of a noisy environment.

\end{abstract}

\maketitle

\section{ Introduction}
 In recent years, quantum information technology has accomplished several realizations. These accomplishments were mainly motivated by the perspectives of quantum computers \cite{nielsen2002quantum, knill2005quantum, ladd2010quantum} and quantum communications \cite{gisin2007quantum, duan2000quantum, monroe2002quantum} that transform and process information encoded in quantum systems. Generally, two encoding schemes are using in quantum information processing. The first one concerns the class of quantum protocols where the information encoding in discrete and finite spectrum systems \cite{bouwmeester2000physics, andersen2015hybrid}. The second class involves quantum information encoded in the states of continuous and infinite spectrum systems \cite{braunstein2005quantum,  weedbrook2012gaussian, ferraro2005gaussian}. The processing of quantum information encoded in continuous variables has attracted considerable attention due to the ease of implementing experimentally, as well as, from the theoretical point of view, it has an elegant mathematical description.
 
Gaussian states are a particular family of continuous-variable systems and are defined by Gaussian Wigner functions \cite{schumaker1986quantum}. This family of states is relatively easy to generate and to manipulated experimentally. Furthermore, from a theoretical viewpoint, it provides benefiting tools to encode and process quantum information with continuous variables due to the limitation of degrees of freedom, which is limited only to the displacement vector and the covariance matrix. Moreover, the Gaussian characteristic of this class of quantum states can be preserved during certain transformations, for instance, the unitary ones such as that associated with the symplectic forms in the phase space \cite{olivares2012quantum, weedbrook2012gaussian}, or non-unitary ones such as those describing the noisy dynamics both Markovian \cite{bellman1957markovian} and non-Markovian \cite{diosi1998non} produced by the inevitable interaction with the environment. Hence, in view to the Gaussian states properties, they have been used successfully to achieve incredible progress over the past decade in several quantum physics areas such as current quantum optical technology \cite{xiao2019continuous, dowling2015quantum, tan2008quantum}, description of optomechanical oscillators \cite{tian2010optical, nunnenkamp2011single}, efficient quantum computation \cite{knill2001scheme}, implementation of quantum teleportation schemes and quantum error correction \cite{olivares2003optimized, wolf2007quantum, ralph2011quantum}. In addition, quantum Gaussian states offer an open and promising technological path, especially in the quantum metrology field \cite{giovannetti2006quantum,gessner2018sensitivity, giovannetti2011advances, paris2009quantum} which has exploited quantum resources to improve the accuracy of measurements. These are essentially motivated by the increased need for more accurate and sensitive detectors.\\
 \hspace*{0.5cm}In the context of improving the sensitivity of detection, quantum metrology has developed strategies based on quantum mechanical resources to achieve the limits of optimal measurements with precision beyond the limits obtained by classical metrology. The main contribution in this field was put by Helstrom \cite{helstrom1976quantum} and Holevo \cite{holevo2011probabilistic},  who established the uncertainty measurement in the estimation protocols of quantum metrology. The first works carried out in quantum metrology were devoted to the single physical parameter estimation. These were extended soon after to the case of estimating several physical parameters. Indeed, some applications require quantum metrological protocols involving several physical unknowns parameters. One may quote for instance; thermometry \cite{correa2015individual, de2016local}, microscopy \cite{hell2009microscopy, hess2006ultra}, super-resolution quantum imaging \cite{tsang2016quantum, yang2016far}, magnetic field detection \cite{nair2016far, zhang2014fitting}, as well as gravitational-wave detection by the use of large interferometers such as VIRGO \cite{acernese2014advanced} and LIGO \cite{abbott2009ligo}. This type of estimation is called multiparameter quantum estimation theory or multiparameter quantum metrology. Due to the incompatible measurement, in multiparameter estimation protocol, we cannot always reach the optimal estimation simultaneously of several parameters encoded in a quantum system. For this, the multiparameter estimation protocol is more challenging than one estimating a single parameter. This difficulty arises from the non-commutativity of the operators associated with the estimated parameters. Hence, the incompatibilities between the optimal measurements \cite{ragy2016compatibility, crowley2014tradeoff, vaneph2013quantum, vidrighin2014joint, proctor2018multiparameter}. Therefore, the quantum limits of precision cannot be attainable generally.\\
 \hspace*{0.5cm}
The importance of Gaussian state formalism in the description of quantum states of light is well established especially, in quantum information theory, including multiparameter quantum metrology. Recently in the literature, there are a certain number of works dedicated to multiparameter quantum estimation theory using the concept of quantum Fisher information matrix (QFIM) together with the Gaussian state formalism \cite{monras2013phase, gao2014bounds, vsafranek2018estimation, carollo2018symmetric, nichols2018multiparameter, bakmou2020multiparameter}. Most of these studies were limited to closed systems and the estimation of the parameters imposed by the unitary transformations. Among these systems, we mention the estimation parameters of Gaussian channels such as; phase change \cite{monras2006optimal, sparaciari2015bounds, sparaciari2016gaussian}, squeezing parameter \cite{milburn1994hyperbolic, chiribella2006optimal, gaiba2009squeezed} and optimal phase for the mode-mixing channel \cite{aspachs2009phase}. Concerning open quantum systems and non-unitary processes, quantum metrology has been extensively applied to finite-dimensional quantum systems as, for instance, in estimating the noise parameter of depolarizing channels \cite{fujiwara2001quantum} or the amplitude damping effect \cite{ji2008parameter}. The quantum metrology based on the continuous variable system remains quite limited, except, some protocols such as the estimation parameters of a lossy bosonic channel \cite{monras2007optimal}, and optimal phase estimation in a Gaussian environment\cite{oh2019optimal}. Also,  it must be noted that the mentioned applications are dedicating to the single parameter estimation.  On the other hand, to the best of our knowledge, very little is known about the use of multiparameter quantum estimation theory in realistic noisy environments. That is the issue that we will address in this work.\\
\vspace{-0.02cm}
\hspace{0.4cm}In this paper, we propose and analyze a measurement scheme of a protocol for simultaneously estimating parameters that are encoding in a displacement operator. The probe state is Gaussian and subjected to an open dynamic. We will investigate the role of quantum entanglement and the purity of the probe Gaussian states to improve the simultaneous precision of measurements. To achieve this purpose, we limit the analysis to an arbitrary two-mode Gaussian probe state that undergoes under a displacement operator which acts only on one of the two modes and subject to a Gaussian noise environment. We start in Sec. \ref{Sec. 2} by offering a brief review of the multiparameter quantum metrology. Next, we present in Sec \ref{Sec. 3} the basic Gaussian concepts used in phase space analysis and their performance in describing the diffusive dynamics of quantum systems in continuous variables. We formulate in Sec. \ref{Sec. 4}, the general framework to study the joint estimation of parameters encoded in the displacement operator acting on the probe Gaussian states and evolving in Gaussian noise environment. In Sec. \ref{Sec. 5}, we investigate the estimation performance over time $t$ for the various state: two modes squeezed vacuum state, two modes squeezed thermal state, two modes coherent state, and two modes coherent thermal state.  Finally, we summarize our work and conclude with some remarks and possible directions for future studies.

\section{Multiparameter quantum metrology}\label{Sec. 2}

A standard scenario of quantum multiparameter estimation theory is to prepare a probe state (the input state)  ${\hat \rho _{inp}}$, which passes through a quantum channel described by the completely positive and trace-preserving linear map ${\Phi  _{\left\{ \boldsymbol{\theta}  \right\}}}$. This map depends on the set of physical parameters one wishes to estimate. The output state is denoted by ${{\hat \rho }_{out}}(\boldsymbol{\theta} ) = {\Phi _{\left\{ \boldsymbol{\theta}  \right\}}}\left( {{{\hat \rho }_{inp}}} \right)$ and then subjected to the measurements to get the values of unknown parameters. Let us consider that a quantum channel depends on the set of parameters $M$. These parameters are building the vector defined in the space parameter named the vector parameter $\boldsymbol{\theta}={\left( {{\theta _1},{\theta _2},...,{\theta _M}} \right)^T}$. The main goal of multiparameter quantum estimation theory is to infer the values of several parameters $\boldsymbol{\theta}$ from the outcomes of generalized measurements $\boldsymbol{x} = {\left( {{x_1},{x_2},...,{x_n}} \right)^T}$ characterized by the POVMs $\left\{\hat \Pi_{\boldsymbol{x}}\right\}$, which are sets of positive operators satisfying ${\hat \Pi _{\boldsymbol{x}}} \ge 0, \quad \int {{\hat \Pi _{\boldsymbol{x}}}d\boldsymbol{x} = \mathbb{1}}$. Furthermore, the joint estimation of the parameters is made using an estimator vector ${\boldsymbol{\theta} ^{est}}\left( \boldsymbol{x} \right)$ which depends on the set of measurement outcomes $\left\{\boldsymbol{x}\right\}$ and satisfies the condition of the unbiased estimator $E\left( {{\boldsymbol{\theta} ^{est}}\left( \boldsymbol{x} \right)} \right) = \sum\limits_{\boldsymbol{x}} {p\left( {\boldsymbol{x} \left| \boldsymbol{\theta}  \right.} \right)} {\boldsymbol{\theta} ^{est}}\left( \boldsymbol{x} \right) = \boldsymbol{\theta}$, where $p\left( {\boldsymbol{x} \left| \boldsymbol{\theta}  \right.} \right) = Tr\left[ {{\hat \Pi _{\boldsymbol{x}}}{{\hat \rho }_{out}}\left( \boldsymbol{\theta}  \right)} \right]$ is the conditional probability distribution to get the value $\boldsymbol{x}$ when the parameters have the values $\boldsymbol{\theta}$. In the several parameters estimation $\boldsymbol{\theta}$, the ultimate precision is fixed by the bounds named quantum Cramér-Rao bounds (QCRBs) \cite{paris2009quantum, braunstein1994statistical, helstrom1967minimum}.

The different QCRBs can be obtained from the QFIM \cite{petz2011introduction,  bakmou2019quantum, paris2009quantum}, which plays an important role in multiparameter quantum metrology. It provides the appropriate tool to get the precision limits for estimating simultaneously several parameters. There are two families of QFIMs extensively used, namely; the symmetric logarithmic derivative (SLD)\cite{helstrom1976quantum, helstrom1969quantum} and right logarithmic derivative (RLD) \cite{yuen1973multiple, fujiwara1995quantum, belavkin1976generalized, fujiwara1994multi, gudder1985holevo, fujiwara1995quantum} QFIMs. Each one has its own QCRBs. The SLD and RLD-QFIMs are defined respectively by
\begin{equation}
F_{{\theta _\mu }{\theta _\nu }}^{\left( S \right)} = Tr\left[ {\hat \rho \frac{1}{2}\left( {\hat L_{{\theta _\mu }}^{\left( S \right)}\hat L_{{\theta _\nu }}^{\left( S \right)} + \hat L_{{\theta _\nu }}^{\left( S \right)}\hat L_{{\theta _\mu }}^{\left( S \right)}} \right)} \right], \label{Eq.1}
\end{equation}
 \begin{equation}
 F_{{\theta _\mu }{\theta _\nu }}^{\left( R \right)} = Tr\left[ {\hat \rho \hat L_{{\theta _\mu }}^{\left( R \right)}\hat L{{_{{\theta _\nu }}^{\left( R \right)}}^\dag }} \right], \label{Eq. 2}
 \end{equation}
 where $\hat L_{{\theta _\mu }}^{\left( S \right)}$ and  $\hat L_{{\theta _\mu }}^{\left( R \right)}$ are the SLD and RLD operators. They satisfy the following differential equations
 \begin{equation}
 {\partial _{{\theta _\mu }}}\hat \rho  = \frac{1}{2}\left( {\hat \rho \hat L_{{\theta _\mu }}^{\left( S \right)} + \hat L_{{\theta _\mu }}^{\left( S \right)}\hat \rho } \right),
 \end{equation}
 \begin{equation}
 {\partial _{{\theta _\mu }}}\hat \rho  = \hat \rho \hat L_{{\theta _\mu }}^{\left( R \right)}.
 \end{equation}
 From the SLD and RLD-QFIMs, one can define two different QCRBs 
 \begin{equation}
 		Tr\left[ {W \mathtt{Cov}\left( {{\boldsymbol{\theta} ^{est}}} \right)} \right] \ge Tr\left[ {{W F^{{{\left( S \right)}^{ - 1}}}}} \right],
 \end{equation}
\begin{equation}
	Tr\left[ {W \mathtt{Cov}\left( {{\boldsymbol{\theta} ^{est}}} \right)} \right] \ge Tr\left[ {{\rm{Re}}{{\left( {W {F^{\left( R \right)}}} \right)}^{ - 1}}} \right] + TrAbs\left[ {{\rm{Im}}{{\left( {{W F^{\left( R \right)}}} \right)}^{ - 1}}} \right],
\end{equation}
where $W$ is a given positive definite matrix allows us to weigh the uncertainty of different estimating parameters. If we choose $W= \mathbb{1}$, we find that the two bounds on the sum of the variances of the estimators of the estimated parameters
 \begin{equation}
n\sum\limits_{\mu  = 1}^M {{\mathop{\rm var}} \left( {{\theta^{est}_\mu }} \right)}  \ge {B_S} = Tr\left[ {{F^{\left( S \right)}}^{ - 1}} \right],\label{5}
 \end{equation}
{\small \begin{equation}
n\sum\limits_{\mu  = 1}^M {{\rm{var}}\left( {{\theta^{est} _\mu }} \right)}  \ge {B_R} = Tr\left[ {{\rm{Re}}{{\left( {{F^{\left( R \right)}}} \right)}^{ - 1}}} \right] + TrAbs\left[ {{\rm{Im}}{{\left( {{F^{\left( R \right)}}} \right)}^{ - 1}}} \right],\label{6}
\end{equation}}
where $TrAbs[A]$ denotes the sum of absolute values of the eigenvalues of a matrix $A$, and $n$ is the number of measurements performed. The SLD-QFIM is the most used to improve the precision of multiparameter quantum estimation via the SLD-QCRB. It is obtained from the Uhlmann’s quantum fidelity between the two output quantum states \cite{braunstein1994statistical}. It has been applied in condensed matter physics to describe criticality and quantum phase transitions \cite{zanardi2007bures, venuti2007quantum, banchi2014quantum, wu2016geometric}, measurement speed limits on the evolution of quantum states \cite{deffner2017quantum, pires2016generalized, del2013quantum, caneva2009optimal}. It has been also applied for the quantification of quantum coherence and quantum entanglement \cite{lu2010quantum, slaoui2019comparative, ye2018quantum, zhong2013fisher, hauke2016measuring}. However, there are various situations of multiparameter quantum estimation protocols in which the RLD-QFIM is given a most informative QCRB \cite{sidhu2019tight, albarelli2020perspective, genoni2013optimal, gao2014bounds}, specifically in protocols connected to the Gaussian formalism. In general, neither the SLD-QCRB nor the RLD-QCRB can be attainable \cite{helstrom1974noncommuting}. Except for the SLD-QCRB, the situation in where the operators associated with the estimated parameters satisfy the following condition \cite{ragy2016compatibility, crowley2014tradeoff, vidrighin2014joint, vaneph2013quantum}
\begin{equation}
		{U_{{\theta _\mu }{\theta _\nu }}} = 0, \quad \text{with} \quad \mu, \nu=1,2,...,M
	\end{equation}
where ${U_{{\theta _\mu }{\theta _\nu }}} =  - \frac{i}{2}Tr\left( {\rho \left[ {\hat L_{{\theta _\mu }}^{\left( S \right)}, \hat L_{{\theta _\nu }}^{\left( S \right)}} \right]} \right)$. This solution is known as the compatibility condition between $\theta_\mu$  and $\theta_\nu$. Hence, ${U_{{\theta _\mu }{\theta _\nu }}}$ is noting as a measure of incompatibility that arises from the inherent quantum nature of the underlying physical system. We stress that this condition involves only the SLD operators because their eigenstates represent the optimal POVM \cite{paris2009quantum, ragy2016compatibility}, in contrast with the optimal measure for RLD operators, which does not always correspond to POVM. In this context and more recently in Ref. \cite{carollo2019quantumness}, was introduced a quantity $\mathcal{R}$ called quantumness. It allows quantifying the deviation of the compatibility condition from the SLD-QCRB and that write as
\begin{equation}
	\mathcal{R} = {\left\| {i {F^{\left( S \right)}}^{ - 1}U} \right\|_\infty },
\end{equation}
where ${\left\| A \right\|_\infty }$ denoted the largest eigenvalue of the matrix $A$. It has been shown that $0 \le \mathcal{R} \le 1$. Thus, the quantumness $\mathcal{R}$ provides a  figure of merit that measures the degree of incompatibility within a multi-parameter estimation protocol. The saturation of the upper bound, $\mathcal{R}=1$, is equivalent to the maximal incompatibility between the measurements of the simultaneously estimated parameters. In the limiting case, $\mathcal{R}=0$, the parameterization model is compatible. 

We remain in the general framework of the multiparameter estimation problem, in which a tighter bound is unified the two bounds $B_R$ and $B_S$ into  Holevo Cramér-Rao bound (HCRB) $B_H$. In a multiparameter model, the HCRB can be expressed by \cite{holevo2011probabilistic, holevo1976noncommutative}
\begin{equation}
	n\sum\limits_\mu ^M {{\mathop{\rm var}} \left( {{\theta^{est} _\mu }} \right)}  \ge {B_H},
\end{equation}
where $B_H$ is defined by
\begin{equation}
{B_H} = \mathop {\min }\limits_{\boldsymbol{\hat X} \in \mathcal{\boldsymbol{\hat X}}} \left\{ {Tr\left[ {{\rm{Re}}{Z_{\boldsymbol{\theta}} }\left[ {\boldsymbol{\hat X}} \right]} \right] + TrAbs\left[ {{\rm{Im}}{Z_{\boldsymbol{\theta}} }\left[ {\boldsymbol{\hat X}} \right]} \right]} \right\}, \label{Eq. 12R2}
\end{equation}
where the minimization is performed on the Hermitian operators that belongings to the set
\begin{equation}
\small	\mathcal{\boldsymbol{\hat X}}=\left\{ {{\hat X_1},{\hat X_2},...,{\hat X_M}\left| {Tr\left[ {\hat \rho {\hat X_\mu }} \right] = 0,Tr\left[ {\left( {{\partial _{{\theta _\mu }}}\hat \rho } \right){\hat X_\nu }} \right]} \right. = {\delta _{\mu \nu }}} \right\},
\end{equation}
and the matrix elements of Hermitian $ M\times M$ matrix $Z_{\boldsymbol{\theta}}$ are defined by
\begin{equation}
	{Z_{{\theta _\mu }{\theta _\nu }}}\left[\boldsymbol{\hat X }\right] = Tr\left[ {\hat \rho {\hat X_\mu }{\hat X_\nu }} \right].
\end{equation}
From Eq. (\ref{Eq. 12R2}), the evaluation of HCRB requires performing an optimization on the sets of Hermitian operators that are not known in general. This makes this optimization very difficult to perform, except for some non-trivial cases \cite{holevo2011probabilistic, bradshaw2017tight, suzuki2019information, yang2019attaining, albarelli2019evaluating}. Recently, it has shown that the HCRB is bounded and satisfy the following inequalities\cite{carollo2019quantumness, albarelli2019upper}
\begin{equation}
	\left( {1 + \mathcal{R}} \right){B_S} \ge {B_H} \ge {B_S}.\label{Eq. 12}
\end{equation}
Since $0\le \mathcal{R} \le 1$, then the estimation through the HCRB is bounded, in fact, by the SLD-QCRB. Therefore, the HCRB cannot provide new information about the scaling of quantum enhancements possible that is not already available in the SLD-QCRB. \\
It is natural to ask which of these bounds is the most informative, i.e. which one is higher and then tighter, depends effectively on the multiparameter estimation model considered. In Refs. \cite{albarelli2020perspective, genoni2013optimal}, the most informative bound is defined by 
\begin{equation}
	 {B_{MI}} = \mathop {\min }\limits_{\left\{ {\text{POVM}\hspace{0.1cm} \hat \Pi } \right\}} \left\{ {Tr\left[ {{I^{ - 1}}} \right]} \right\},
\end{equation}
where $I$ is the classical Fisher information matrix that is given by
\begin{equation}
	{I_{{\theta _\mu }{\theta _\nu }}} = \sum\limits_{\boldsymbol{x}} {\left( {\frac{{\partial \log \,p\left( {\left. \boldsymbol{x} \right|\boldsymbol{\theta} } \right)}}{{\partial {\theta _\mu }}}\frac{{\partial \log p\left( {\left. \boldsymbol{x} \right|\boldsymbol{\theta} } \right)}}{{\partial {\theta _\nu }}}} \right)} p\left( {\left. \boldsymbol{x} \right|\boldsymbol{\theta} } \right).
\end{equation}
The most informative bound satisfies the following inequalities \cite{albarelli2020perspective}
\begin{equation}
	n \sum\limits_\mu ^M {{\mathop{\rm var}} \left( {{\theta^{est}_\mu }} \right)}  \ge {B_{MI}} \ge {B_H} \ge \max \left\{ {{B_R},{B_S}} \right\}.
\end{equation}
From Eq.(\ref{Eq. 12}), the maximum incompatibility condition, $\mathcal{R}=1$, implied that the SLD-QCRB gives an estimate of the HCRB up to a factor of 2. In other words, the upper bound that is possible attainable by HCRB is $B_H^{max}=B_S\left(1+\mathcal{R}\right)$, which becomes $2B_S$ in the maximal incompatible condition and coincides with $B_S$ in the asymptotic limit. Therefore, we have the following tight bound of HCRB
\begin{equation}
	2{B_S}\ge B_S\left(1+\mathcal{R}\right) \ge {B_H} \ge \max \left\{ {{B_R},{B_S}} \right\}. \label{EQ. 17}
\end{equation} 
By evaluating the upper bound of HCRB, this inequality allows us to predict the behavior of HCRB.
\section{ Basic concepts of Gaussian states}\label{Sec. 3}
We consider a canonical infinite-dimensional system composed of $m$-bosonic modes, each mode $k$ is described by a pair of quadrature field operators $\hat {q}_k$, $\hat{p}_k$ acting on a Hilbert space $\mathcal{H}_k$. The space $\mathcal{H}_k$ is spanned by a number basis $\left\{ {{{\left| n \right\rangle }_k}} \right\}$ of eigenstates of the number operator $\hat{n}_{k}=\hat a_k^\dag {\hat a_k}$. The quadrature operators can be expressed in terms of the ladder operators $\hat a_k^\dag$ and $\hat a_k$ as ${\hat q_k} = {{{\hat a}_k} + \hat a_k^ + }$,   ${\hat p_k} = i\left( {\hat a_k^ +  - {{\hat a}_k}} \right)$. In the phase-space description of $m$-modes, the quadrature operators $\hat {q}_k$ and $\hat{p}_k$ are collected in a vector $\mathbf{\hat R}= {\left( {{{\hat q}_1},{{\hat p}_1},...,{{\hat q}_m},{{\hat p}_m}} \right)^T}$. The canonical commutation relations between the operators can be written in compact form with natural units ($\hbar=2$)
\begin{equation}
\left[ {{{\hat R}_j},{{\hat R}_k}} \right] = 2i \hspace{0.1cm}{\Omega _{jk}},
\end{equation}
where ${\Omega _{jk}}$ are the elements of the symplectic matrix $\Omega$ of dimension $2m \times 2m$
\begin{equation}
\Omega  = \mathop  \oplus \limits_{k = 1}^m \omega,   \hspace{1.5cm}  \omega  = \left[ {\begin{array}{*{20}{c}}
	0&1\\
	{ - 1}&0
	\end{array}} \right].
\end{equation}

In the context of quantum information processing, the information has encoded in a quantum state represented by the matrix density $\hat \rho$ living in a Hilbert space $\mathcal{H}$, and it is a Hermitian positive semi-definite matrix usually normalized, i.e. $Tr\left[\hat \rho\right]=1$. In the case in which the quantum information is encoded in the states living in a finite Hilbert space, we have the discrete variable system. In the opposite case, if the matrix density lived in the Hilbert space of infinite-dimensional, thus we talk about the continuous variables system described by observables with continuous eigenspectra. In the last case, the quantum archetype is represented by $m$-bosonic modes corresponding to $m$-quantum harmonic oscillators described in phase-space. It is important to emphasize that any density matrix can be represented in terms of the quasi-probability distribution defined over phase-space. This representation is called the Wigner representation that is characterized by the characteristic function $\chi_{\hat \rho } \left( \boldsymbol{R} \right)$ or Wigner function
\begin{equation}
{\chi _{\hat \rho }}\left( {\mathbf{R}} \right) = Tr\left[ {{{\hat D}_{ - {\mathbf{R}}}} \hspace{0.1cm}\hat \rho } \right],
\end{equation}
where $\mathbf{R} = {\left( {{q_1},{p_1},...,{q_m},{p_m}} \right)^T}$ is a vector of $2m$ real coordinates in phase-space and ${\hat D_{ - {\boldsymbol{R}}}}$ is the Weyl operator which is defined by 
\begin{equation}
{\hat D_{ - {\mathbf{R}}}} = {e^{ - i{{\mathbf{R}}^T}\Omega {\mathbf{\hat R}}}}.
\end{equation}
In what follows, we consider the Gaussian states whose Wigner representation (${\chi _{\hat \rho }}$ or ${\mathcal{W}_{\hat \rho }}$) is Gaussian, i.e.
\begin{equation}
{\chi _{\hat \rho }}\left( \mathbf{R} \right) = exp\left( { - \frac{1}{2}{\mathbf{R}^T}\left( {\Omega \sigma {\Omega ^T}} \right)\mathbf{R} - i{{\left( {\Omega \left\langle {\mathbf{\hat R}} \right\rangle } \right)}^T}\mathbf{R}} \right)
\end{equation}
The Wigner representation of the Gaussian states is completely described only by two important statistical quantities, which are the first and second moments. The first moment called displacement vector and  expressed by
\begin{equation}
 \mathbf{d}=\left\langle {\mathbf{\hat R}} \right\rangle=Tr\left[ {\hat \rho \hspace{0.1cm}\mathbf{\hat R}} \right], \label{13}
\end{equation}
and the second moment is the covariance matrix $\sigma$ which given by
\begin{equation}
{\sigma _{jk}} = \frac{1}{2}\left\langle {{{\hat R}_j}{{\hat R}_k} + {{\hat R}_k}{{\hat R}_j}} \right\rangle  - \left\langle {{{\hat R}_j}} \right\rangle \left\langle {{{\hat R}_k}} \right\rangle. \label{Eq. 24}
\end{equation}
 The covariance matrix $\sigma$ is a $2m \times 2m$ real symmetric matrix defined strictly positive and satisfy the  following uncertainty inequality \cite{simon1987gaussian, simon1994quantum}
\begin{equation}
\sigma  + i \hspace{0.1cm}\Omega  \ge 0. \label{15}
\end{equation}
The evolution of a quantum state is describing as a typical transformation called a quantum channel.  This quantum channel is characterizing by a linear map $\Phi_{\left\{ \boldsymbol{\theta}  \right\}} :\hat \rho  \to \Phi_{\left\{ \boldsymbol{\theta}  \right\}}\left( {\hat \rho } \right)$, which is completely positive and trace-preserving $Tr\left[ {\Phi_{\left\{ \boldsymbol{\theta}  \right\}} \left( {\hat \rho } \right)} \right]=1$. The simplest quantum channels are the ones represented by unitary transformations $\hat U {\hat U^\dag } = \mathbb{1}$, and the state of a quantum system evolved  according to the following rule
\begin{equation}
	{{\hat \rho }} \to  \hat U{{\hat \rho }}{\hat U^\dag }.
\end{equation}
When the unitary transformations preserve the Gaussian characteristic of the initial quantum state, we say that these transformations are Gaussian unitary transformations. They are generated via $\hat U = \exp \left( { - i\hat H} \right)$, with $\hat H$ represents the Hamiltonians of the system. These are second-order polynomials in the field operators (creation and annihilation operators). The set of transformations coming from these Hamiltonians individualizes the type of unitary Gaussian operations. Each Gaussian unitary transformation corresponds to a symplectic transformation in the phase space, i.e. a linear transformation preserving the symplectic form
\begin{equation}
\hat S\Omega {\hat S^T} = \Omega ,
\end{equation}
where $\hat S$ is a $2m \times 2m$ real symplectic matrix. Under this symplectic transformation, the statistical moments  of Eq. (\ref{13}) and Eq. (\ref{Eq. 24}) transformed as
\begin{equation}
{\mathbf{d}} \to \hat S{\mathbf{d}} + {{\mathbf{R}}}, \hspace{0.5cm}{\sigma } \to \hat S\hspace{0.1cm}{\sigma }{\hat S^\dag}.
\end{equation}

Noisy Gaussian channels are generally subject to noise and loss of quantum coherence due to interaction with the environment. The dynamics of the $m$-mode bosonic system coupled to an environment evolving under a noisy Gaussian non-unitary channel are describing by the following master equation
 \begin{equation}
\small \dot {\hat \rho}  = \frac{\gamma }{2}\sum\limits_{k = 1}^m {\left( {\left( {{N_e} + 1} \right){\cal L}\left[ {{{\hat a}_k}} \right] + {N_e}{\cal L}\left[ {\hat a_k^\dag } \right] - {{\bar M}_e}{\cal D}\left[ {{{\hat a}_k}} \right] - {M_e}{\cal D}\left[ {\hat a_k^\dag } \right]} \right)\hat \rho }, \label{19}
\end{equation}
where the dot denotes the time derivative, $\gamma$  is the overall damping rate, while $N_e \in \mathbb{R}$ and $M_e \in \mathbb{C}$ represents the effective photon numbers and the squeezing parameter  of the baths respectively. The positivity of the density matrix $\hat \rho$ imposes the important constraint ${\left| M_e \right|^2} \le N_e\left( {N_e + 1} \right)$. For $M_e=0$ the bath is at thermal equilibrium and $N_e$ coincides with the average number of thermal photons, i.e. $N_e=\bar{n}$. The Lindblad super-operators in Eq. (\ref{19}) are defined by $\mathcal{L}\left[ {\hat O} \right]\hat \rho  = 2\hat O\hat \rho {\hat O^\dag } - {\hat O^\dag }\hat O\hat \rho  - \hat \rho {\hat O^\dag }\hat O$ and $\mathcal{D}\left[ {\hat O} \right]\hat \rho  = 2\hat O\hat \rho \hat O - \hat O\hat O\hat \rho  - \hat \rho \hat O\hat O$. The terms involving $\mathcal{L}\left[ {\hat a} \right]$ and $\mathcal{L}\left[ {\hat a^\dag} \right]$ describe losses and linear, phase-insensitive. The terms proportional to $\mathcal{D}\left[\hat{a}\right]$ and $\mathcal{D}\left[\hat a^\dag\right]$ describe phase dependent fluctuations. The considered $m$-mode input Gaussian state with ${\hat \rho _{inp}}$ of the first moment and the second moment are respectively denoted by $\mathbf{d}_{inp}$ and $\sigma_{inp}$. We note that the time evolution governed by the master equation preserves the Gaussian characteristic, and the dissipation action on the first and second moments at time $t$ are given by \cite{ferraro2005gaussian}
 \begin{equation}
 \sigma \left( t \right) = {e^{ - \gamma  t}}\sigma_{inp} + \left( {1 - {e^{ - \gamma  t}}} \right){\sigma _\infty }, \hspace*{0.4cm}\mathbf{d}\left( t \right) = {e^{\frac{{ - \gamma t}}{2}}}\mathbf{d}_{inp},  \label{20}
 \end{equation}
where ${\sigma _\infty } = \mathop  \oplus \limits_{k = 1}^m {\sigma _{k,\infty }}$ is the diffusion matrix which is expressed  in terms of the bath parameters as follows
\begin{equation}
{\sigma _{k,\infty }} = \left[ {\begin{array}{*{20}{c}}
	{1 + 2N_e + 2{\mathop{\rm Re}\nolimits} \left[ M_e \right]}&{2{\mathop{\rm Im}\nolimits} \left[ M_e \right]}\\
	{2{\mathop{\rm Im}\nolimits} \left[ M_e \right]}&{1 + 2N_e - 2{\mathop{\rm Re}\nolimits} \left[ M_e \right]}
	\end{array}} \right].
\end{equation}
In particular, the second moment at any given time $t$ Eq.(\ref{20}) must be satisfying the usual uncertainty relations of Eq. (\ref{15}).
\vspace{0.2cm}
\section{General framework}\label{Sec. 4}
In this section, we formalize a framework to study the simultaneous estimation of two parameters $\theta_1$ and  $\theta_2$ encoded in a Gaussian channel described by the displacement operator $\hat D\left( {{\theta _1},{\theta _2}} \right) = \exp \left( {i {\mathbf{\theta _2}}\hat q - i{\mathbf{\theta _1}}\hat p} \right)$, which acts only on one of the two-mode of a Gaussian probe state. The realization of this study will be performing by preparing the initial probe Gaussian state $\hat \rho_{inp}$. One of two-mode is displacing by the acting of the displacement operator  $\hat D\left( {{\theta _1},{\theta _2}} \right)$. After this displacement, the two modes are subject to the effects of a Gaussian environment. The latter can be considered as an unavoidable part of the detection procedure. The state of the system, after the interaction with the environment, is given by $\hat \rho_{out}$. The two modes of the output state are mixing with a beam splitter, and the Gaussian homodyne measurement is performing. To illustrate this metrological scheme,  we give the Fig. (\ref{Fig1}). We mention that the precision carried out on the estimated parameters depends on the output states.

\begin{figure}
\tikzset{every picture/.style={line width=0.75pt}} %set default line width to 0.75pt        

\begin{tikzpicture}[x=0.44pt,y=0.7pt,yscale=-1,xscale=1]
	%uncomment if require: \path (0,326); %set diagram left start at 0, and has height of 326
	
	%Shape: Ellipse [id:dp8177003503627258] 
	\draw  [fill={rgb, 255:red, 155; green, 155; blue, 155 }  ,fill opacity=0.75 ] (49,170) .. controls (49,120.29) and (57.73,80) .. (68.5,80) .. controls (79.27,80) and (88,120.29) .. (88,170) .. controls (88,219.71) and (79.27,260) .. (68.5,260) .. controls (57.73,260) and (49,219.71) .. (49,170) -- cycle ;
	%Straight Lines [id:da8664056934202531] 
	\draw    (81,99) -- (80,100) -- (125,99) ;
	%Shape: Rectangle [id:dp870788339342879] 
	\draw  [fill={rgb, 255:red, 184; green, 233; blue, 134 }  ,fill opacity=0.76 ] (123,76) -- (196,76) -- (196,130) -- (123,130) -- cycle ;
	%Straight Lines [id:da15472714142919686] 
	\draw    (198,101) -- (234,101) -- (245,101) ;
	%Shape: Ellipse [id:dp6383880012605088] 
	\draw  [fill={rgb, 255:red, 128; green, 128; blue, 128 }  ,fill opacity=0.7 ] (232,165) .. controls (232,109.77) and (258.86,65) .. (292,65) .. controls (325.14,65) and (352,109.77) .. (352,165) .. controls (352,220.23) and (325.14,265) .. (292,265) .. controls (258.86,265) and (232,220.23) .. (232,165) -- cycle ;
	%Straight Lines [id:da9967321915028318] 
	\draw    (406,100) -- (436,100) ;
	%Straight Lines [id:da9646497713665036] 
	\draw    (406,239) -- (431,238) ;
	%Straight Lines [id:da5855226181692006] 
	\draw [color={rgb, 255:red, 74; green, 144; blue, 226 }  ,draw opacity=0.98 ][line width=3]    (463,168) -- (497,168) ;
	%Straight Lines [id:da1827385707982656] 
	\draw    (436,100) -- (478,166) ;
	%Straight Lines [id:da6016595478964735] 
	\draw    (433,238) -- (480,168) ;
	%Straight Lines [id:da8142075555019361] 
	\draw    (480,166) -- (518,102) ;
	%Straight Lines [id:da8582716476367585] 
	\draw    (518,102) -- (544,102) ;
	%Straight Lines [id:da3307446229625459] 
	\draw    (518,242) -- (540.5,242) ;
	%Flowchart: Delay [id:dp0463067427829682] 
	\draw  [fill={rgb, 255:red, 155; green, 155; blue, 155 }  ,fill opacity=0.75 ] (540,79) -- (562.5,79) .. controls (574.93,79) and (585,119.74) .. (585,170) .. controls (585,220.26) and (574.93,261) .. (562.5,261) -- (540,261) -- cycle ;
	%Shape: Ellipse [id:dp7257142455474563] 
	\draw  [color={rgb, 255:red, 0; green, 0; blue, 0 }  ,draw opacity=1 ][fill={rgb, 255:red, 155; green, 155; blue, 155 }  ,fill opacity=0.75 ] (372,169.5) .. controls (372,119.52) and (381.63,79) .. (393.5,79) .. controls (405.37,79) and (415,119.52) .. (415,169.5) .. controls (415,219.48) and (405.37,260) .. (393.5,260) .. controls (381.63,260) and (372,219.48) .. (372,169.5) -- cycle ;
	%Straight Lines [id:da8352876543880441] 
	\draw    (333,239) -- (380,239) ;
	%Straight Lines [id:da15448733471440002] 
	\draw    (339,100) -- (379,99) ;
	%Straight Lines [id:da25149515720172433] 
	\draw    (80,242) -- (253,242) ;
	%Straight Lines [id:da531572284239934] 
	\draw    (479,168) -- (520,242) ;
	
	% Text Node
	\draw (131,80) node [anchor=north west][inner sep=0.85pt] [rotate=-32.09] {$\hat D(\theta_1,\theta_2)$};
	% Text Node
	\draw (422,161) node [anchor=north west][inner sep=0.75pt]    {$BS$};
	% Text Node
	\draw (547.37,228.48) node [anchor=north west][inner sep=0.75pt]  [rotate=-270.1] [align=left] {Homodyne detection};
	% Text Node
	\draw (235,136) node [anchor=north west][inner sep=0.75pt]   [align=left] {\textbf{ \ Gaussian}\\\textbf{ \ \ \ thermal }\\\textbf{envirnoment}};
	% Text Node
	\draw (57.51,179.79) node [anchor=north west][inner sep=0.75pt]  [rotate=-271.09]  {$\hat\rho_{inp}$};
	% Text Node
	\draw (380.53,179.79) node [anchor=north west][inner sep=0.75pt]  [rotate=-271.03]  {$\hat \rho_{out}$};

\end{tikzpicture}

	\centering
	
	\captionsetup{justification=centerlast, singlelinecheck=false}\captionof{figure}{Schematic to illustrate the adaptive protocol for estimating the displacement parameters under the noises Gaussian environment. Initially, we prepare the probe state $\hat \rho_{inp}$, using the various Gaussian operations as illustrated in Sec. (\ref{Sec. 4}). Next, the probe state is submitting to an unknown displacement $\hat D\left(\theta_1, \theta_2\right)$. After this displacement, the two-mode is evolving under a Gaussian thermal environment. The two modes of output state are mixing with a beam splitter (BS), and then the homodyne detection measurement is performed.}
	\label{Fig1}
\end{figure}
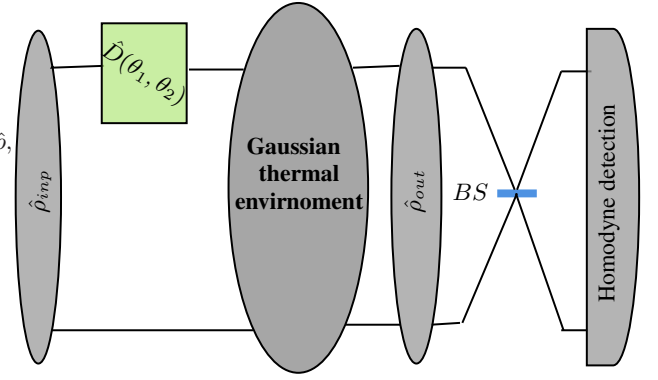

In general, to the best of our knowledge, there is a gap between the estimation precision achieved by homodyne detection measurement and that achieved by evaluating SLD, RLD-QCRBs. That is not excluding since the SLD and RLD-QCRBs are not tights. In what follows, we addressed the following questions: From the probe state displaced by the action of $\hat D\left( {{\theta _1},{\theta _2}} \right)$, and in the noise environment, how can we estimate simultaneously, with precision, the pair of parameters?  In other words, in the presence of a noise environment, can we evaluate tight bounds of estimation precision? And can we attain an high estimation precision than that obtained by HD measurement\cite{genoni2017cramer}? In this work, the input state considering is a general two-mode squeezed displaced thermal state, which is given by 
\begin{equation}
{\hat \rho _{inp}} = {\hat S_2}\left( \xi  \right)\hat D\left( \alpha  \right)\left( {{\hat \rho _{th}} \otimes {\hat \rho _{th}}} \right)\hat D{\left( \alpha  \right)^\dag }{\hat S_2}{\left( \xi  \right)^\dag }, \label{Eq.22}
\end{equation}
where ${\hat S_2}\left( \xi  \right) = \exp \left( {\xi {{\hat a_1}^\dag }{{\hat a_2}^\dag } - {\xi ^ * }\hat a_1\hat a_2} \right)$ is the two-mode squeezing operator with the squeezed parameter $r$ and the rotation angle $\phi$ ($\xi=r e^{\phi}$), and $\hat D\left( \alpha  \right) = \exp \left( {{\alpha _1}\hat a_1^\dag  - \alpha _1^*{{\hat a}_1} + {\alpha _2}\hat a_2^\dag  - \alpha _2^*{{\hat a}_2}} \right)$ is the two-mode displacement operator, with $ \alpha_{k=(1,2)}=q_k+ip_k$ is the parameter of coherent light, and $\hat \rho_{th}$ denote the thermal states, which is given by 
	\begin{equation}
		{\hat \rho _{th}} = \sum\limits_n {\frac{{{{\bar n}^n}}}{{{{\left( {\bar n + 1} \right)}^{n+1}}}}\left| n \right\rangle } \left\langle n \right|, \label{Eqq. 33}
	\end{equation}
	where ${\bar n} = \left\langle {\hat a^\dag {\hat a}} \right\rangle= Tr\left(\hat \rho_{th}\hat a^\dag {\hat a} \right)$  is the mean number of photons in the bosonic mode, which is expressed in terms of the temperature effect as ${\bar n} = {\left( {{e^\beta } - 1} \right)^{ - 1}}$. In the limit of zero temperature, one recovers the two-modes pure vacuum state $\left| 0 0 \right\rangle \left\langle 0 0 \right|$.

\vspace*{-0.3cm}
\section{results and discussion}\label{Sec. 5}
\vspace{-0.2cm}
In this part,  we focus on two kinds of probe Gaussian states;  the first, Gaussian pure states, which, in turn, decomposed into two situations that  derived from Eq. (\ref{Eq.22});  (i) two-modes squeezed vacuum state $\hat \rho _{inp}^{\left( i \right)} = {\hat S_2}\left( \xi  \right)\left| {00} \right\rangle \left\langle {00} \right|{\hat S_2}{\left( \xi  \right)^\dag }$, which derived from Eq. (\ref{Eq.22}) with $\alpha=0$ and $\bar n=0$, and (ii) two-modes displacement vacuum state also called two-modes coherent state $\hat \rho _{inp}^{\left( {ii} \right)} = \hat D\left( \alpha  \right)\left| {00} \right\rangle \left\langle {00} \right|\hat D{\left( \alpha  \right)^\dag }$, which derived from Eq. (\ref{Eq.22}) with $\xi  = 0$ and $\bar n=0$. The second kind, namely Gaussian mixed states, which decomposed into two categories; (iii)  two-modes squeezing thermal state $\hat \rho _{inp}^{\left( iii \right)} = {\hat S_2}\left( \xi  \right)\left( {{\hat \rho _{th}} \otimes {\hat \rho _{th}}} \right){\hat S_2}{\left( \xi  \right)^\dag }$,  that derived from Eq. (\ref{Eq.22}) with $\alpha=0$, (iiii)  two-modes coherent thermal state $\hat \rho _{inp}^{\left( iiii \right)} = \hat D\left( \alpha  \right)\left( {{\hat \rho _{th}} \otimes {\hat \rho _{th}}} \right)\hat D{\left( \alpha  \right)^\dag }$, that derived from Eq. (\ref{Eq.22})  with $\xi=0$. These classes of states are typical states that can be implemented and used in quantum information protocols with continuous-variable systems. For example,  in optomechanical systems, the coherent states provide the appropriate description of laser-produced states \cite{tian2010optical, nunnenkamp2011single}. While the squeezed states present non-classic characteristics such as the potential ability to generate quantum entanglement \cite{wolf2004gaussian, wolf2003entangling}, which is an important resource for performing the diverse protocols in different disciplines of quantum information theory.\\
To simplify our purpose, we assume that the two modes are identical, and the squeezing parameter of baths takes the value zero ($M_e=0$). This assumption means that the photon number of the Gaussian environment corresponds to the photon number of thermal states, i.e. $N_e= \bar{n}$.
\vspace{-0.5cm}
\subsection{Gaussian pure states}
 Let us now consider two types of Gaussian probe pure states:  two-modes squeezed vacuum state and two-modes displaced vacuum state. We start with the two-modes squeezed vacuum state
 \vspace{-0.5cm}
  \subsubsection{Two-modes squeezed vacuum state (TMSV)}
 We will be trying to examine the case when the two-modes squeezed vacuum (TMSV) state,  also known as Einstein-Podolski-Rosen (EPR) state,  is considered as a probe state.  In this case, the input state of Eq. (\ref{Eq.22}) reduced to
  \begin{equation}
  	{\hat \rho _{inp}} = {\hat S_2}\left( \xi  \right)\left| {00} \right\rangle \langle 00|{\hat S_2}{\left( \xi  \right)^\dag },
  \end{equation}
where $ {\hat S_2}\left( \xi  \right)\left| {00} \right\rangle$  is expressed in Fock space by
\begin{equation}
 {\hat S_2}\left( \xi  \right)\left| {00} \right\rangle  = \frac{1}{{\cosh r}}{\sum\limits_{n = 0} {\left( { - {e^{i\phi }}\tanh r} \right)} ^n}\left| {n,n} \right\rangle.
\end{equation}
For every two-modes squeezing  $r > 0 $, we have the EPR correlations between the quadrature, which implies that the two-mode are entangled. In the Heisenberg picture, the evolution of two modes, by the action of the squeezing operator, is describing by the Bogoliubov transformations, which are given by
\begin{equation}
{\hat S_2}\left( \xi  \right){\hat a_1}{\hat S_2}{\left( \xi  \right)^\dag } = \cosh\left( r \right){\hat a_1} + {e^{i\varphi }}\sinh \left( r \right){\hat a_2}^\dag, 
\end{equation}
\begin{equation}
{\hat S_2}\left( \xi  \right){\hat a_2}{\hat S_2}{\left( \xi  \right)^\dag } = \cosh\left( r \right){\hat a_2} - {e^{i\varphi }}\sinh \left( r \right){\hat a_1}^\dag.
\end{equation}
The mean photon number of the TMSV state is given by $\left\langle {{{\hat a}_1}^\dag {{\hat a}_1}} \right\rangle  = \left\langle {{{\hat a}_2}^\dag {{\hat a}_2}} \right\rangle  = \sinh {\left( r \right)^2}$. The first and second moment of this input state are given by $\mathbf{d}_{inp}=0$, and
\begin{equation}
{\sigma _{inp}} = \left( {\begin{array}{*{20}{c}}
	{\cosh \left( {2r} \right){\mathbb{1}_{2 \times 2}}}&{\sinh \left( {2r} \right){\hat R_{\varphi}}}\\
	{\sinh \left( {2r} \right){\hat R_{\varphi}}}&{\cosh \left( {2r} \right){\mathbb{1}_{2 \times 2}}}
	\end{array}} \right),
\end{equation} 
where $\hat R_{\varphi}$ is the symplectic transformation related to the squeezing angle. It is defined by
\begin{equation}
{\hat R_\varphi } = \left( {\begin{array}{*{20}{c}}
	{\cos \varphi }&{\sin \varphi }\\
	{\sin \varphi }&{ - \cos \varphi }
	\end{array}} \right).
\end{equation}
After the action  of the displacement operator only on one of the two-modes, the evolution of the TMSV state at a time $t$  is  characterized by ${\mathbf{d}_{out}}\left( t \right)  = {e^{ - \frac{{\gamma t}}{2}}}{\left( {{\theta _1},{\theta _2},0,0} \right)^T}$ and
\begin{equation}
\sigma _{out}\left( t \right) = {e^{ - \gamma t}}{\sigma _{inp}} + \left( {1 - {e^{ - \gamma t}}} \right)\left( {1 + 2{N_e}} \right){1_{4 \times 4}},
\end{equation}
By using the result of Refs. \cite{gao2014bounds, bakmou2020multiparameter}, the two QCRBs from Eqs. (\ref{5}) and (\ref{6})  can be immediately evaluated as

	\begin{widetext}
		
\begin{equation}
{B_S}\left( t \right) = \left( {{{\rm{e}}^{t\gamma }} - 1} \right)\left( {1 + 2 N_e} \right) + \cosh \left( {2r} \right) - \frac{{\sinh {{\left( {2r} \right)}^2}}}{{\left( {{{\rm{e}}^{t\gamma }} - 1} \right)\left( {1 + 2 N_e} \right) + \cosh \left( {2r} \right)}}, \label{Eq. 43}
\end{equation}
 \begin{equation}
 {B_R}\left( t \right) = 2{{\rm{e}}^{t\gamma }}\left( {1 + N_e} \right) + \cosh \left( {2r} \right) - \left( {2 N_e + 1} \right) - \frac{{\sinh {{\left( {2r} \right)}^2}}}{{2\left( {{{\rm{e}}^{t\gamma }} - 1} \right) N_e + \cosh \left( {2r} \right) - 1}}.
\end{equation}

\end{widetext}
Now we consider a general Gaussian measurement described by a covariance matrix $\sigma_{Gm}$ satisfied $\sigma_{Gm}+i \Omega \ge  0$. For a single-mode Gaussian state, the most general Gaussian measurement corresponds to a displaced squeezed vacuum state, which is characterizing by  
\begin{equation}
		{\sigma _{Gm}}(s,\phi ) = \hat R(\phi )\left( {\begin{array}{*{20}{c}}
				{{e^{ - 2s}}}&0\\
				0&{{e^{2s}}}
		\end{array}} \right)\hat R{(\phi )^\dag },
	\end{equation} 	
	with $\hat R\left(\phi\right)=\left( {\begin{array}{*{20}{c}}
			{\cos \phi }&{\sin \phi }\\
			{ - \sin \phi }&{\cos \phi }
	\end{array}} \right)$. The limits $s \to \infty$ and $s \to 0$ describe, respectively, the homodyne detection and the heterodyne detection. In the case taking into account the inefficient detection that comes by the effect of a noisy environment, the Gaussian measurement matrix is described by
\begin{equation}
	\sigma _{Gm}^{({\rm{ineff }})} = X{\sigma _{Gm}}{X^\dag } + {Y^\dag },
\end{equation}
where $X= {e^{\frac{{\gamma t}}{2}}}\mathbb{1}_{2\times 2}$ and $Y = \left( {{e^{\gamma t}} - 1} \right)\mathbb{1}_{2\times 2}$. It easily to general this result in the case of two-modes Gaussian measurements, such that the overall measurement matrix $\sigma_{Gm}^{(\rm{ineff})}$ is obtained by taking the direct sum of the single-mode measurements, i.e. ${\sigma _{Gm}}^{({\rm{ineff }})} = {\sigma _{Gm\left( {{a_1}} \right)}}^{({\rm{ineff }})} \oplus {\sigma _{Gm\left( {{a_2}}\right)}}^{({\rm{ineff }})}$. By exploiting the result of Ref. \cite{genoni2017cramer}, for TMSV, the CR bound evaluated via homodyne detection is given by
\begin{widetext}
	\begin{equation}
		{B_{HD}}\left(t\right) = 2N_e\left( {{{\rm{e}}^{t\gamma }} - 1} \right) + {{\rm{e}}^{2t\gamma }}\left( {1 + \cos {{\left( \phi  \right)}^2}} \right) + \cosh \left( {2r} \right) - \frac{{\sinh {{\left( {2r} \right)}^2}}}{{2N_e\left( {{{\rm{e}}^{t\gamma }} - 1} \right) + {{\rm{e}}^{2t\gamma }}\left( {1 + \cos {{\left( \phi  \right)}^2}} \right) + \cosh \left( {2r} \right) - 1}}.
	\end{equation}
\end{widetext}
To determine the degree of incompatibility between the parameters of this model, we will calculate the quantumness parameter $\mathcal{R}$, which one can obtain as
\begin{equation}
	\mathcal{R} = \frac{{{{\rm{e}}^{t\gamma }}}}{{{{\rm{e}}^{t\gamma }} - 2{N_e}\left( {1 - {{\rm{e}}^{t\gamma }}} \right) + \cosh \left( {2r} \right) - 1}}. \label{Eq. 480}
\end{equation}
When ${N_e} \to \infty$, the quantumness $\mathcal{R}$ tends to $0$, which means that the multiparameter model is compatible and the indeterminacy that arises from the quantum nature of the system disappears. This limit is known as the asymptotic limit, and the multiparameter model is called the asymptotically classical-quantum statistical model \cite{albarelli2020perspective}. In addition, the performance of any measurement is very poor and does not support the ultimate estimation precision. In this limit, we have $B_H=B_S$.\\
In order to compute the upper bound of HCRB, one can be exploiting the results of Eq. (\ref{Eq. 43}) and Eq. (\ref{Eq. 480})  to find the $B_H^{max}$, that has expressed as
\begin{widetext}
\begin{equation}
		B_H^{\max }\left(t\right) = \left( {1 + \frac{{{{\rm{e}}^{t\gamma }}}}{{\left( {{{\rm{e}}^{t\gamma }} - 1} \right)\left( {1 + 2{N_e}} \right) + \cosh \left( {2r} \right)}}} \right)\left( {\left( {{{\rm{e}}^{t\gamma }} - 1} \right)\left( {1 + 2{N_e}} \right) + \cosh \left( {2r} \right) - \frac{{\sinh {{\left( {2r} \right)}^2}}}{{\left( {{{\rm{e}}^{t\gamma }} - 1} \right)\left( {1 + 2{N_e}} \right) + \cosh \left( {2r} \right)}}} \right)
\end{equation}
\begin{figure}[H]
		\centering
		\begin{subfigure}{.33\textwidth}
			\centering
			\includegraphics[width=5.9cm]{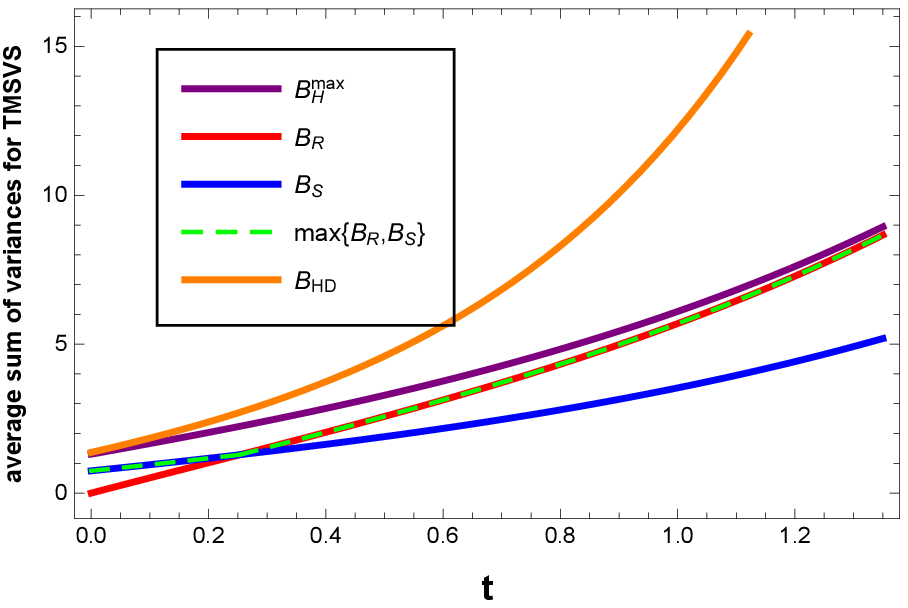}
			\caption{}\label{F2a}
		\end{subfigure}
		\begin{subfigure}{.33\textwidth}
			\centering
			\includegraphics[width=5.9cm]{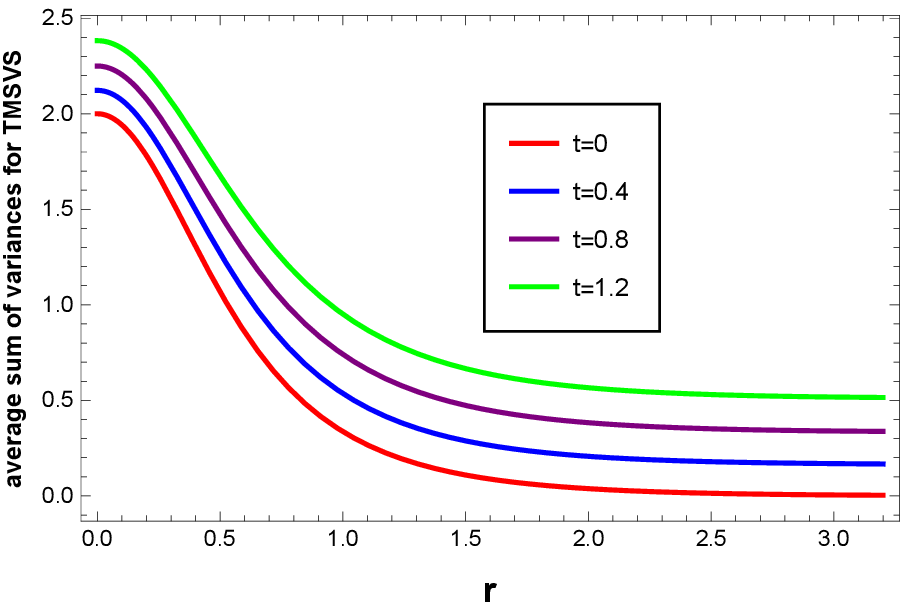}
			\caption{}\label{F2b}
		\end{subfigure}
		\begin{subfigure}{.33\textwidth}
			\centering
			\includegraphics[width=5.9cm]{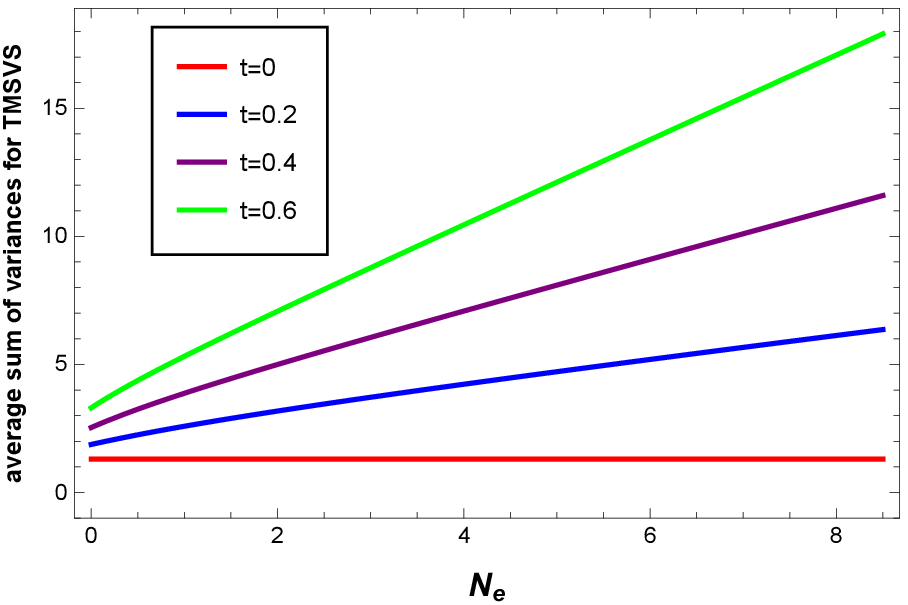}
			\caption{}\label{F2c}
		\end{subfigure}
		\renewcommand\thefigure{\arabic{figure}}
		\addtocounter{figure}{-1}
		\captionsetup{justification=raggedright, singlelinecheck=false, labelfont=sc} \captionof{figure}{The plot of the average sum of variances for the two-modes squeezed vacuum probe state. Fig. (\ref{F2a}) represents the SLD, RLD-QCRBs, the max of the SLD and RLD-QCRBs, the upper bound of HCRB, and the HD bound as functions of time $t$,  with $N_e= 0.5$  is the mean number of a photon of the Gaussian environment,  and the initial squeezing parameter of the squeezed vacuum state is taking equal to $r=0.4$, the homodyne angle of the beam-splitter is fixed in $\pi/4$. Fig. (\ref{F2b}) represents the upper bound of HCRB as a function of the squeezed parameter for different values of time  $t$ when the mean number of photons of the Gaussian environment $N_e=0.5$. Fig. (\ref{F2c}) represents the upper bound of HCRB as a function of the mean number of photon $N_e$ for different values of time $t$ when the initial squeezing parameter takes the value $0.4$ and the overall damping rate $\gamma=1$.}\label{Fig2}
	\end{figure}

\end{widetext}
 Fig. (\ref{Fig2}) shows the variation of the sum of mean square error as a function of the different parameters of the system. Fig. (\ref{F2a}) showing the behavior of SLD, RLD-QCRBs, and the max of SLD and RLD-QCRBs, the upper bound of HCRB, and the CR bound evaluated via HD, as functions of the interaction time $t$ with the environment. The SLD, RLD-QCRBs, the max of SLD and RLD-QCRBs, the upper bound of HCRB, and the CR bound evaluated via HD are increasing functions of the interaction time $t$ with the Gaussian environment. From Fig. (\ref{F2a}) in shorter interaction time, we noted that the upper bound of HCRB is greater than the SLD, RLD-QCRBs, and coincides with HD bound, which implied that the upper bound of HCRB is tightening by HD bound. While in more extended interaction time, i.e. in the limits of larger times, the upper bound of HCRB turned out to be approximately the RLD-QCRB. This means that the parameter model, in the the limits of larger times, will be a D-invariant quantum statistical model, in which finding that $B_H=B_R$. If we compared the performance attained by the upper bound of HCRB, and which gained by homodyne detection,  in the short interaction time, i.e. in the small values of times,  we find that the sum of the mean squared error obtained from the homodyne detection is corresponding to the upper bound of HCRB, which indicated that the upper bound of HCRB is tight and that the homodyne detection is the ultimate measurement. But in in the limits of larger times,  the upper bound of HCRB renders a better performance than which obtained by homodyne detection measurement. Fig. (\ref{F2b}) shows the behavior of the upper bound of HCRB, which is a decreasing function of the initial squeezing parameter $r$ and increasing function of the interaction time $t$.  We notice that the ultimate precision of measurement in the estimated parameters (minimal values of the upper bound of HCRB) has been adapting when $t = 0$ (Red solid line), which corresponds to the first control that was performing before the interaction with the environment. Moreover, it is remarkable to notice that once the TMSV state is undergoing a displacement under the effect of a Gaussian thermal environment, the upper bound of HCRB starts to increase, which implied that we are beginning to lose the ultimate precision that has been obtaining in $t=0$, this due to the effect of the thermal environment. It does not mean that the ultimate accuracy is completely disappeared since, with the great values of the squeezed parameter, the upper bound of HCRB still reach the minimum even under the noise environment. One can explain this by the entangled nature of the TMSV state. Thus, when the input state become entangled, the measurement accuracy is affected imperceptibly by the effect of noise.  From the behavior represented in Fig. (\ref{F2c}), we notice that the upper bound of HCRB evolves rapidly when the average number of Gaussian thermal environment photons $N_e$ takes the great values. It means that the increase of $N_e$ is not suitable for estimating the unknown displacement parameters that describe the displacement of TMSV state under the noise effect.
 \vspace{-0.8cm}
 \subsubsection{Two-modes displacement vacuum state (TMDV)}
 Now, we are going to examine the case in where the input state is the two-modes coherent state. Thus, Eq. (\ref{Eq.22}) reduced to 
\begin{equation}
	 \hat \rho _{inp} = \hat D\left( \alpha  \right)\left| {00} \right\rangle \left\langle {00} \right|\hat D{\left( \alpha  \right)^\dag },  \label{32}
\end{equation}
 where $\hat D\left( \alpha  \right)\left| {00} \right\rangle$ is the displacement vacuum state, that is expressed in Fock space by \cite{chai1992two}
\begin{equation}
\left| {{\alpha _1},{\alpha _2}} \right\rangle  = {e^{ - \frac{{{{\left| {{\alpha _1}} \right|}^2}}}{2}}}{e^{ - \frac{{{{\left| {{\alpha _2}} \right|}^2}}}{2}}}\sum\limits_{{n_1},{n_2} = 0} {\frac{{{\alpha _1}^{{n_1}}{\alpha _2}^{{n_2}}}}{{\sqrt {{n_1}!{n_2}!} }}} \left| {{n_1},{n_2}} \right\rangle,
\end{equation}
where ${\alpha _k}$ is the parameter of coherent light. Two mode coherent states $\left| {{\alpha _1},{\alpha _2}} \right\rangle$ are the eigenstates of the  annihilation operators $\hat{a}_k$, i.e. 
${{\hat a}_k}\left| {{\alpha _1},{\alpha _2}} \right\rangle  = {\alpha _k}\left| {{\alpha _1},{\alpha _2}} \right\rangle$ and satisfy the following relation
\begin{equation}
\frac{1}{\pi^2 }\int {\left| {{\alpha _1},{\alpha _2}} \right\rangle \left\langle {{\alpha _1},{\alpha _2}} \right|} d{\alpha _1}d{\alpha _2} = 1.
\end{equation}
In the Heisenberg picture, the annihilation operator is transforming by the linear unitary Bogoliubov transformation
\begin{equation}
\hat D\left( {{\alpha _k}} \right){{\hat a}_k}\hat D{\left( {{\alpha _k}} \right)^\dag } = {{\hat a}_k} + {\alpha _k}. \label{Eq.34}
\end{equation}
Using Eq. (\ref{Eq.34}), one can express the translation of the quadrature operators as $\boldsymbol{\hat R} \to \boldsymbol{\hat R }+ \boldsymbol{R}$ where $\boldsymbol{\hat R} = {\left( {{{\hat q}_1},{{\hat p}_1},{{\hat q}_2},{{\hat p}_2}} \right)^T}$ and $\boldsymbol{R} = {\left( {{q_1},{p_1},{q_2},{p_2}} \right)^T}$. The mean photon number of TMDV state can be expressed as $\left\langle {\hat a_1^\dag {{\hat a}_1}} \right\rangle  = \left\langle {\hat a_2^\dag {{\hat a}_2}} \right\rangle  = {\left| {{\alpha _1}} \right|^2} = {\left| {{\alpha _2}} \right|^2}$. The first and second moments of the input state from Eq. (\ref{32}) are given by
\begin{equation}
{\mathbf{d}_{inp}} = {\left( {{q_1},{p_1},{q_2},{p_2}} \right)^T} \hspace{0.7cm};  \hspace{0.7cm} {\sigma _{inp}} = {\mathbb{1}_{4 \times 4}}.
\end{equation}
After the action of displacement operator only on one mode and the interaction with the Gaussian environment, the output state is characterized by  ${\bf{d}}_{out}\left( t \right) = {e^{ - \frac{{\gamma t}}{2}}}{\left( {{q_1} + {\theta _1},{p_1} + {\theta _2},{q_2},{p_2}} \right)^T}$ and
\begin{equation}
\sigma _{out}\left( t \right) = {e^{ - \gamma t}}{\mathbb{1}_{4 \times 4}} + \left( {1 - {e^{ - \gamma t}}} \right)\left( {1 + 2{N_e}} \right){\mathbb{1}_{4 \times 4}}.
\end{equation}
 In this case, the two QCRBs $B_S$ and $B_R$ of Eqs. (\ref{5}) and Eq. (\ref{6}) can be straightforwardly obtained as
\begin{equation}
{B_S}\left( t \right) = {{\rm{e}}^{t\gamma }}\left( {1 + 2 N_e} \right) - 2 N_e, \label{Eq.37}
\end{equation}
\begin{equation}
{B_R}\left( t \right) = {{\rm{e}}^{t\gamma }}\left( {1 + 2 N_e} \right) + {{\rm{e}}^{t\gamma }} - 2 N_e. \label{Eq.38}
\end{equation}
By exploiting the results of Ref. \cite{genoni2017cramer},  one can be quickly evaluated the CR bound via homodyne detection measurement as 
	\begin{equation}
		{B_{HD}}\left( t \right) = 2N_e\left( {{{\rm{e}}^{t\gamma }} - 1} \right) + \frac{1}{2}{{\rm{e}}^{2t\gamma }}\left( {3 + \cos \left( {2\phi } \right)} \right). \label{EQ. 49}
\end{equation}
Similarly, the calculation of the quantumness parameter $\mathcal{R}$, in this case, leads to
	\begin{equation}
		\mathcal{R} = \frac{{{{\rm{e}}^{t\gamma }}}}{{{{\rm{e}}^{t\gamma }} - 2 N_e\left( {1 - {{\rm{e}}^{t\gamma }}} \right)}}. \label{Eq. 59}
	\end{equation}
When $N_e \to 0$, we have $\mathcal{R}\to 1$. Thus, the incompatibility reaches, in approximately, the maximum possible. By using the results of Eq. (\ref{Eq.37}) and Eq. (\ref{Eq. 59}), one gets that the upper bound of HCRB as
	\begin{equation}
		B_H^{\max }\left(t\right) = {{\rm{e}}^{t\gamma }}\left( {1 + 2{N_e}} \right) +{\rm{e}}^{t\gamma}- 2{N_e}, \label{EQQ60}
	\end{equation}
	which exactly corresponds to the RLD-QCRB of Eq. (\ref{Eq.38}) this means that $B_H= B_R$ and therefore the model, in this case, is D-invariant.\\
Fig. (\ref{Fig3}) represents the average sum of variances for the two-mode coherent probe state. Fig. (\ref{F3a}) describes the SLD, RLD-QCRBs, the max of SLD and RLD-QCRBs, the HCRB that coincides, in this case, with the RLD-QCRB, and the bound evaluate via HD as functions of the interaction time $t$.  In short values of interaction time $t$, we noted that the HCRB coincides with the bound evaluated by HD. Whereas, in large values of time $t$, the behavior of HCRB varied under the HD bound, which means that the HD measurement is the ultimate measurement and HCRB is tightly bound. While Fig. (\ref{F3b}) shows the HCRB as increasing functions of the average number of Gaussian thermal environment photons $N_e$. In general, we observe that the HCRB and the HD bound are both increases with increasing the interaction time $t$ and $N_e$. Thus, the two modes coherent probe state is not appropriate for estimating the displacement parameters in the presence of environmental noise. This can be explained by the equations (\ref{Eq.37},\ref{Eq.38},\ref{EQ. 49}, \ref{EQQ60}) in which the bounds do not depend on the average energy of the coherent probe state.
\begin{widetext}
	
	\begin{figure}[H]
	\centering
	\begin{subfigure}{.49\textwidth}
		\centering
		\includegraphics[width=8.5cm]{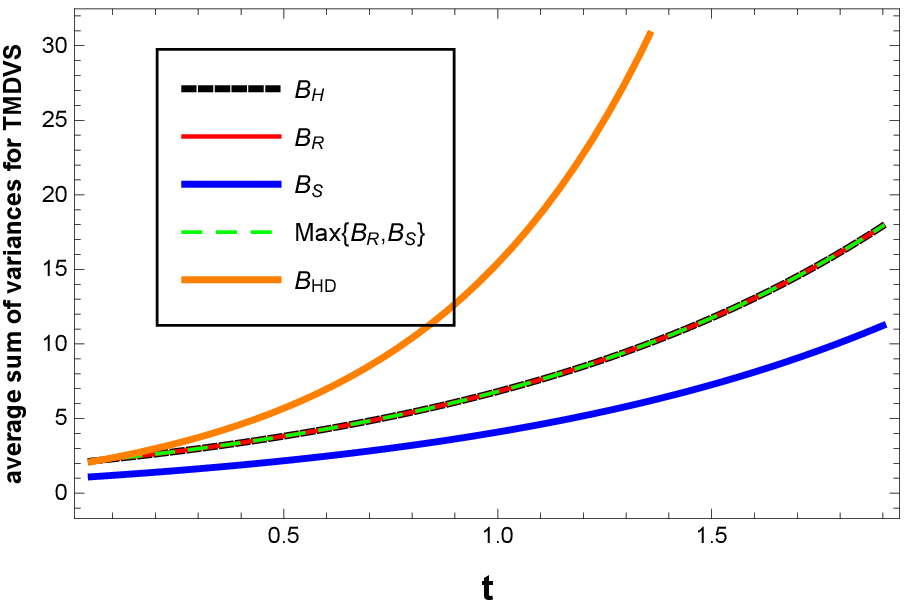}
		\caption{}\label{F3a}
	\end{subfigure}
	\begin{subfigure}{.49\textwidth}
		\centering
		\includegraphics[width=8.5cm]{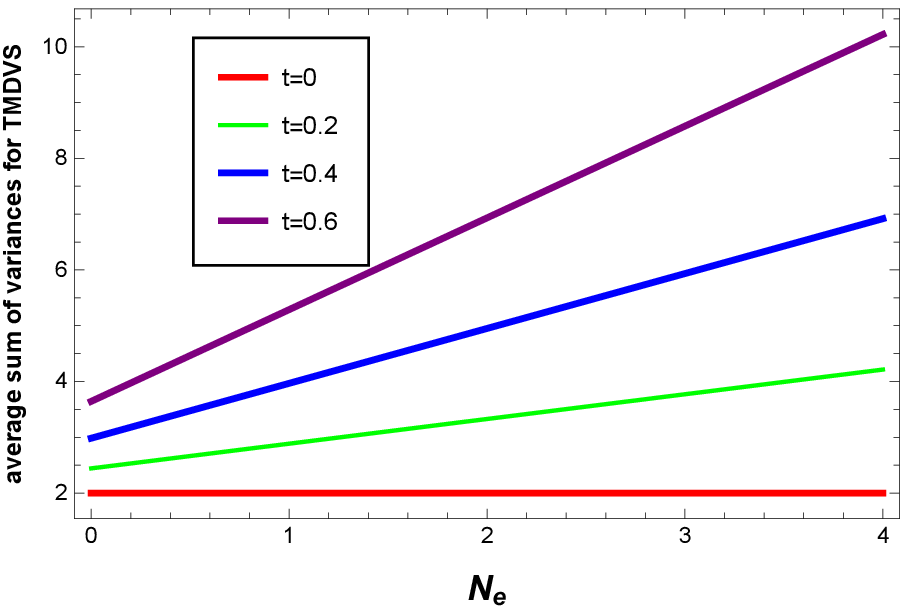}
		\caption{}\label{F3b}
	\end{subfigure}
	\renewcommand\thefigure{\arabic{figure}}
	\addtocounter{figure}{-1}
	\captionsetup{justification=raggedright, singlelinecheck=false, labelfont=sc} \captionof{figure}{The plot of the average sum of variances for the two-modes displacement vacuum probe state. Fig. (\ref{F3a}) represents the SLD, RLD-QCRBs, the max of SLD and RLD-QCRBs, and the HCRB( which coincides with the RLD-QCRB in this case), and the HD bound as functions of the interaction time $t$,  with the mean number of a photon of the Gaussian environment $N_e= 0.5$,  and the homodyne angle of the beam-splitter fixed in $\pi/4$.  Fig. (\ref{F3b}) represents the HCRB as a function of the mean number of photon $N_e$ for different values of time $t$ and the overall damping rate $\gamma=1$.}\label{Fig3}
\end{figure}
\vspace{-0.5cm}
\end{widetext}
\subsection{Gaussian mixed states}
Let us consider the more general case, in where the input states are mixed.  We will focus on two of the most interesting cases:  the first concerns a two-modes squeezed thermal state, and the second case corresponds to a two-modes coherent thermal state. In both cases, we consider that the displacement operator characterized by two unknown parameters is acting only on one of the two modes.
\vspace{-0.7cm}
\subsubsection{Two-modes squeezed thermal state (TMST)}
In this case, the input state of Eq. (\ref{Eq.22}) reduced to
\begin{equation}
{\hat \rho _{inp}} = {\hat S_2}\left( \xi  \right)\left( {{\hat \rho _{th}} \otimes {\hat \rho _{th}}} \right){\hat S_2}{\left( \xi  \right)^\dag }, \label{Eq.39}
\end{equation}
where $\hat \rho_{th}$ is the thermal state that given in Eq. (\ref{Eqq. 33}). The first and second moment for this input state reads as $\mathbf{d}_{inp}=0$ and
 \begin{equation}
 {\sigma _{inp}} = (2\bar{n} + 1)\left( {\begin{array}{*{20}{c}}
 	{\cosh \left( {2r} \right){\mathbb{1}_{2 \times 2}}}&{\sinh \left( {2r} \right){\hat R_\varphi }}\\
 	{\sinh \left( {2r} \right){\hat R_\varphi }}&{\cosh {{\left( {2r} \right)}\mathbb{1}_{2 \times 2}}}
 	\end{array}} \right).
 \end{equation}
 The average number of photons in TMST is $\left\langle {\hat a_k^\dag {{\hat a}_k}} \right\rangle = {\sinh ^2}r + \bar{n}$ (for $k=1,2$). One can describe the TMST under interaction with the Gaussian environment by the following first and second moments. ${\mathbf{d}_{out}}\left( t \right)  = {e^{ - \frac{{\gamma t}}{2}}}{\left( {{\theta _1},{\theta _2},0,0} \right)^T}$ and
\begin{equation}
\sigma _{out}\left( t \right) = {e^{ - \gamma t}}(2\bar{n} + 1){\sigma _{inp}} + \left( {1 - {e^{ - \gamma t}}} \right)\left( {1 + 2{N_e}} \right){\mathbb{1}_{4 \times 4}}.
 \end{equation}
  The SLD and RLD-QCRBs are computed as
 \begin{equation}
 {B_S}\left( t \right) = \frac{{\left( {1 + 2 N_e} \right)\left( {2 - 2{{\rm{e}}^{t\gamma }} + {{\rm{e}}^{2t\gamma }} + 2\left( {{{\rm{e}}^{t\gamma }} - 1} \right)\cosh \left( {2r} \right)} \right)}}{{\cosh \left( {2r} \right) + {{\rm{e}}^{t\gamma }} - 1}}, \label{Eq. 64}
 \end{equation} 
 \begin{equation}
 {B_R}\left( t \right) = \frac{{2{N_e}\left( {1 + {N_e}} \right){{\rm{e}}^{2t\gamma }} + 2\left( {{{\rm{e}}^{t\gamma }} - 1} \right){{\left( {1 + 2{N_e}} \right)}^2}\sinh {{\left( r \right)}^2}}}{{{N_e}{{\rm{e}}^{t\gamma }} + \left( {1 + 2{N_e}} \right)\sinh {{\left( r \right)}^2}}}.
 \end{equation}
By using Ref. \cite{genoni2017cramer}, one can readily evaluate the CR bound via homodyne detection as
\begin{widetext}
\begin{equation}	
	{B_{HD}}\left( t \right) = 2N_e{{\rm{e}}^{t\gamma }} + \frac{1}{2}{{\rm{e}}^{2t\gamma }}\left( {3 + \cos \left( {2\phi } \right)} \right) + \left( {2 + 4N_e} \right)\sinh {\left( r \right)^2} - \frac{{2{{\left( {1 + 2N_e} \right)}^2}\sinh {{\left( {2r} \right)}^2}}}{{4N_e{{\rm{e}}^{t\gamma }} + {{\rm{e}}^{2t\gamma }}\left( {3 + \cos \left( {2\phi } \right)} \right) + \left( {4 + 8N_e} \right)\sinh {{\left( r \right)}^2}}}.
\end{equation}
\end{widetext}

In this situation, the quantumness parameter $\mathcal{R}$ that measure the incompatibility in a multiparameter model can compute as 
\begin{equation}
	\mathcal{R} = \frac{{{{\rm{e}}^{t\gamma }}}}{{\left( {1 + 2N_e} \right)\left( { + {{\rm{e}}^{t\gamma }} + \cosh \left( {2r} \right) - 1} \right)}}. \label{Eq. 67}
\end{equation}
In the limit of ${N_e} \to \infty $, we have $\mathcal{R} \to 0$. Therefore, from Eq. (\ref{Eq. 12}), one can conclude approximately that $B_H=B_S$, this means that the model is asymptotically classical. In this limit, the performance of any measurement is very powerless and does not support the ultimate estimation precision. By using the results of Eq. (\ref{Eq. 64}) and Eq. (\ref{Eq. 67}), one can evaluate the upper bound of HCRB as

\begin{widetext}
\begin{equation}
	B_H^{\max }\left( t \right) = \frac{{\left( {2 - 2{{\rm{e}}^{t\gamma }} + {{\rm{e}}^{2t\gamma }} + 2\left( {{{\rm{e}}^{t\gamma }} - 1} \right)\cosh \left( {2r} \right)} \right)\left( {2{{\rm{e}}^{t\gamma }}\left( {1 + {N_e}} \right) + \left( {2 + 4{N_e}} \right)\sinh {{\left( {2r} \right)}^2}} \right)}}{{{{\left( {{{\rm{e}}^{t\gamma }} + \cosh \left( {2r} \right) - 1} \right)}^2}}}
	\end{equation}
	\begin{figure}[H]
		\centering
		\begin{subfigure}{.33\textwidth}
			\centering
			\includegraphics[width=5.95cm]{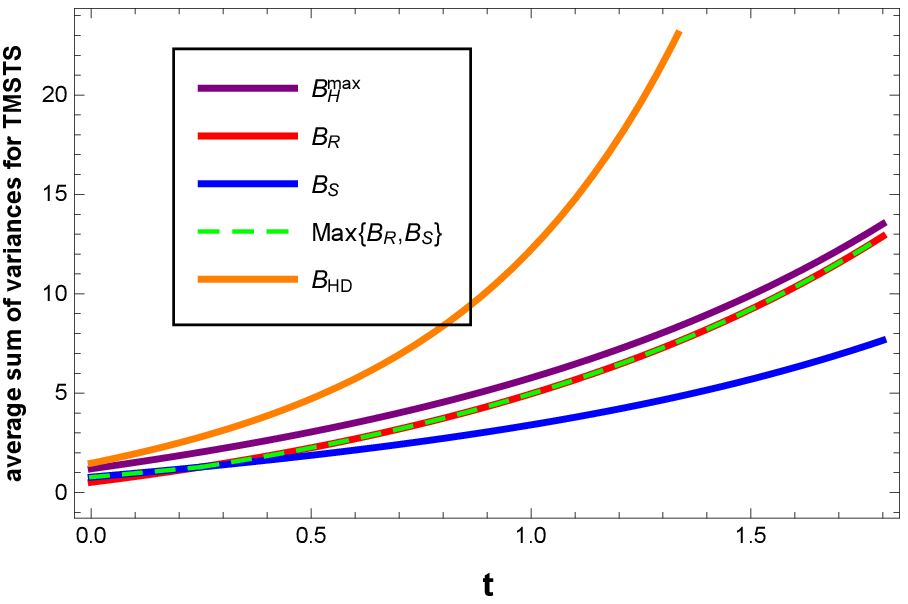}
			\caption{}\label{F4a}
		\end{subfigure}
		\begin{subfigure}{.33\textwidth}
			\centering
			\includegraphics[width=5.95cm]{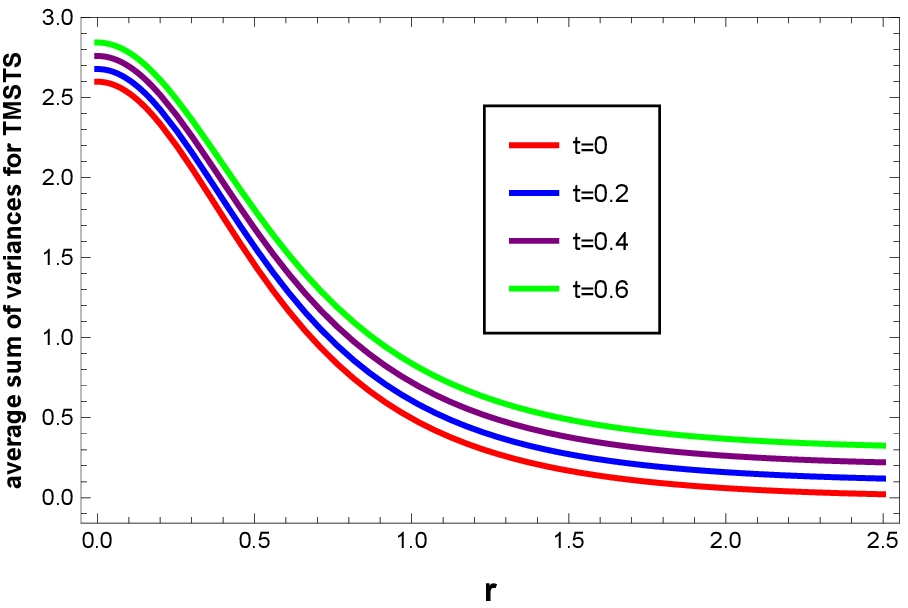}
			\caption{}\label{F4b}
		\end{subfigure}
		\begin{subfigure}{.33\textwidth}
			\centering
			\includegraphics[width=5.95cm]{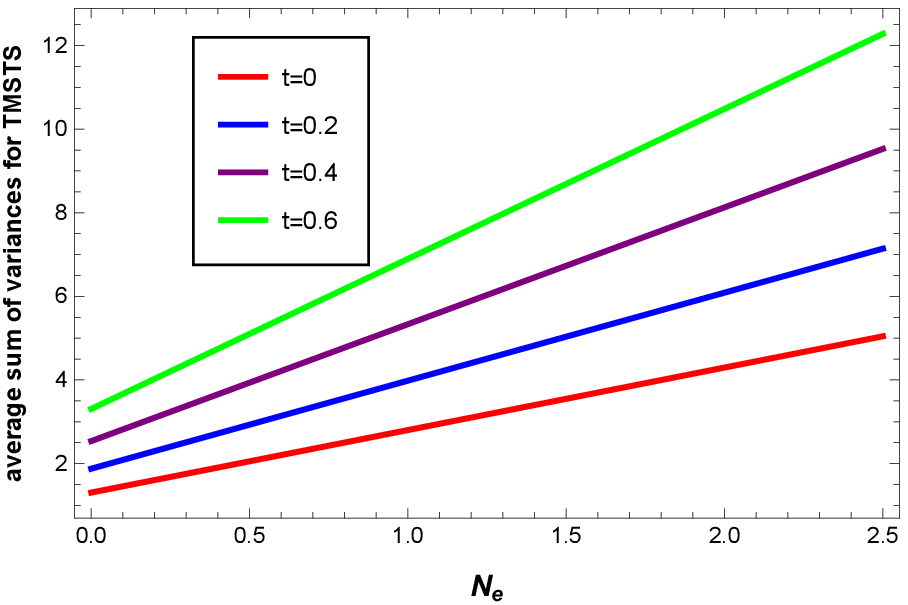}
			\caption{}\label{F4c}
		\end{subfigure}
		\renewcommand\thefigure{\arabic{figure}}
		\addtocounter{figure}{-1}
		\captionsetup{justification=raggedright, singlelinecheck=false, labelfont=sc} \captionof{figure}{The plot of the average sum of variances for the two-modes squeezed thermal probe state. Fig. (\ref{F4a}) represents the SLD, RLD-QCRBs, the max of SLD and RLD-QCRBs, the upper bound of HCRB, and HD bound as functions of the interaction time $t$, with $N_e= 0.5$ is the mean number of a photon of the Gaussian environment, and the initial squeezing parameter of the squeezed thermal state takes $r=0.4$, and the homodyne angle of beam-splitter has fixed in $\pi/4$. Fig. (\ref{F4b}) represents the upper bound of HCRB as a function of the squeezed parameter for different values of time $t$ when the mean number of photons of the Gaussian environment $N_e=0.5$. Fig. (\ref{F4c}) represents the upper bound of HCRB as a function of the mean number of photon $N_e$ for different values of time $t$ when the initial squeezing parameter takes the value $0.4$, and the overall damping rate $\gamma=1$.}\label{Fig4}
	\end{figure}
\end{widetext}
The results obtained, when considering the mixed squeezed thermal states as probe states, are reported in Fig. (\ref{Fig4}). Fig. (\ref{F4a}) represents SLD, RLD-QCRBs, the max of SLD and RLD-QCRBs, the upper bound of HCRB and HD bound as increasing functions in terms of the interaction time $t$. From Fig. (\ref{F4a}), in the limit of small values of time t, the upper bound of HCRB is larger than the SLD, RLD-QCRBs, and coincide with the HD bound. While for greats values of $t$, the upper bound of HCRB turns out to be approximately the RLD bound, i.e. in the large values of times and from Eq. (\ref{Eq. 12}), we have $B_{H}^{max}=B_R=B_H$, thus the parameter model becomes D-invariant. On the other hand, in short time $t$, we discover that homodyne detection is the ultimate measurement, and the upper bound of HCRB is, in approximately, tight bound; because we find that the sum of the mean squared error obtained from the homodyne detection is corresponding to the upper bound of HCRB. But in large values of time $t$, the upper bound of HCRB renders a better performance than which obtained by homodyne detection measurement. Fig. (\ref{F4b}) shows the behaviors of the upper bound of HCRB as a decreasing function of $r$ for the different values of $t$. We set the average number of thermal photon at $\bar{n}=N_e=0.5$. From Fig. (\ref{F4b}),  we observe that the upper bound of HCRB has decreased and reached the minimum values when $r$ takes large values. It means that the accuracy enhancement has been achieving when the TMST state becomes more entangled. In addition, the ultimate measurement has been obtaining during the first control at $t=0$, which was performing before the interaction with the environment (Red solid line). However, the performance of the measurements results is not nasty when there is noise imposed by the environment, as the upper bound of HCRB remains minimal provided that $r$ becomes large. These results are similar to those obtained for the pure TMSV probe state. Fig. (\ref{F4c}) represents the behaviors of the upper bound of HCRB as a function of the average number of thermal photons, which equal to the average number of Gaussian thermal environment photons ($\bar{n}=N_e$), for different values of interaction time $t$. We observe that the upper bound of HCRB increases rapidly when $N_e$ increases. It implies that an increase in $N_e$ is not a good choice for the simultaneous estimation of the displacement parameters.
\vspace{-0.5cm}
 \subsubsection{ Two-modes displacement thermal state (TMDT)}
Now, we consider the two-modes mixed coherent thermal state as a probe state. It is also called two modes displaced thermal (TMDT) state, which is obtained from the general state of Eq. (\ref{Eq.22}) and given by
 \begin{equation}
 {\hat \rho _{inp}} = \hat D\left( \alpha  \right)\left( {{\hat\rho _{th}} \otimes {\hat \rho _{th}}} \right)\hat D{\left( \alpha \right)^\dag }. \label{Eq.48}
 \end{equation}
 This probe state described by the following first and second moments
${\mathbf{d}_{inp}} = 2{\left( {{\mathop{\rm Re}\nolimits} \left[ {{\alpha _1}} \right],{\mathop{\rm Im}\nolimits} \left[ {{\alpha _1}} \right],{\mathop{\rm Re}\nolimits} \left[ {{\alpha _2}} \right],{\mathop{\rm Im}\nolimits} \left[ {{\alpha _2}} \right]} \right)^T}$ and
\begin{equation}
{\sigma _{inp}} = \left( {2\bar{n} + 1} \right){\mathbb{1}_{4 \times 4}}.
\end{equation}
The average number of photons in TMDT is $\left\langle {\hat a_k^\dag {{\hat a}_k}} \right\rangle  = {\left| \alpha  \right|^2} + \bar{n}$ (for $k=1, 2$). The first and second moments of the input state (\ref{Eq.48}) where each of their modes is coupled to an independent Gaussian thermal environment with loss rate $\gamma$ and number of thermal excitation $N_e$ obey the diffusion equation
 \begin{equation}
  \mathbf{d}_{out}\left( t \right)  = 2{e^{ - \frac{{\gamma t}}{2}}}{\left( {{\mathop{\rm Re}\nolimits} \left[ {{\alpha _1}} \right] + {\theta _1},{\mathop{\rm Im}\nolimits} \left[ {{\alpha _1}} \right] + {\theta _2},{\mathop{\rm Re}\nolimits} \left[ {{\alpha _2}} \right],{\mathop{\rm Im}\nolimits} \left[ {{\alpha _2}} \right]} \right)^T}
  \end{equation}
  \begin{equation}
\sigma _{out}\left( t \right) = {e^{ - \gamma t}}(2\bar{n} + 1) + \left( {1 - {e^{ - \gamma t}}} \right)\left( {1 + 2{N_e}} \right){1_{4 \times 4}},
\end{equation}
where $\bar{n}=N_e$ corresponds to the average number photon of Gaussian thermal environment. The SLD, and RLD-QCRBs can be evaluating from Eqs. (\ref{5}) and (\ref{6}). They are given by
\begin{equation}
{B_S}\left( t \right) = {{\rm{e}}^{t\gamma }}\left( {1 + 2 N_e} \right), \label{Eq. 47}
\end{equation}
\begin{equation}
{B_R}\left( t \right) = {{\rm{e}}^{t\gamma }}\left( {1 + 2 N_e} \right) + {{\rm{e}}^{t\gamma }}. \label{Eq. 48}
\end{equation}
By using the results of Ref. \cite{genoni2017cramer}, one can evaluate the CR bound via homodyne detection as
	\begin{equation}
		{B_{HD}}\left( t \right) = 2{N_e}{{\rm{e}}^{t\gamma }} + \frac{1}{2}{{\rm{e}}^{2t\gamma }}\left( {3 + \cos \left( {2\phi } \right)} \right). \label{EQQ73}
	\end{equation}
	In this situation, the quantumness parameter of incompatibility is given by
\begin{equation}
		\mathcal{R} = \frac{1}{{1 + 2N_e}}.\label{REQ. 73}
\end{equation}
When $N_e \to 0$, we have $\mathcal{R}=1$. Thus, the model is maximal incompatible. In addition, we noted that, from Eq. (\ref{REQ. 73}), the quantumness parameter $\mathcal{R}$ does not depend on the value of time $t$.  Hence, the nature of the multiparameter model does not change over time $t$. By applying the results of Eq. (\ref{Eq. 47}) and Eq. (\ref{REQ. 73}), one can evaluate the upper bound of HCRB as follows;
\begin{equation}
	{B_{H}^{max}}\left( t \right) = {{\rm{e}}^{t\gamma }}\left( {1 + 2 N_e} \right) + {{\rm{e}}^{t\gamma }}, \label{Eq. 777}
\end{equation}
which coincides with the RLD-QCRB that has given in Eq. (\ref{Eq. 48}) this implied that $B_H^{max}=B_R=B_H$ and therefore the model is D-invariant statistical model.
 \begin{widetext}
 	
 	\begin{figure}[H]
 		\centering
 		\begin{subfigure}{.49\textwidth}
 			\centering
 			\includegraphics[width=8cm]{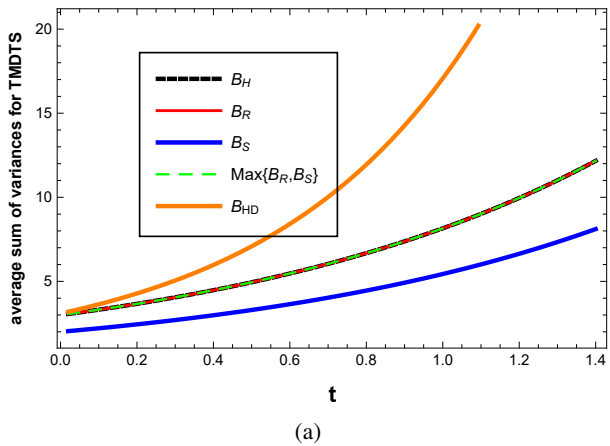}
 			\caption{}\label{F5a}
 		\end{subfigure}
 		\begin{subfigure}{.49\textwidth}
 			\centering
 			\includegraphics[width=8cm]{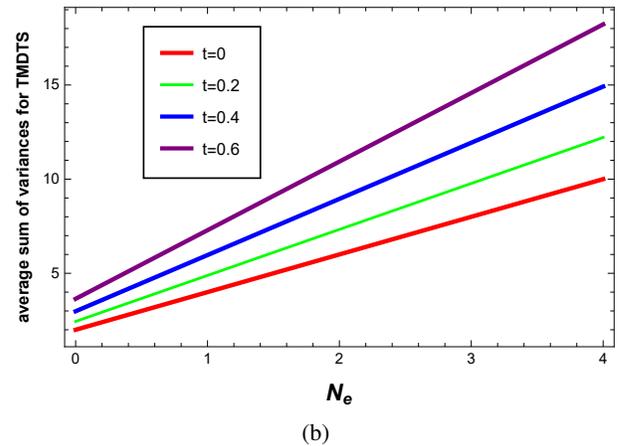}
 			\caption{}\label{F5b}
 		\end{subfigure}
 		\renewcommand\thefigure{\arabic{figure}}
 		\addtocounter{figure}{-1}
 		\captionsetup{justification=raggedright, singlelinecheck=false, labelfont=sc} \captionof{figure}{The plot of the average sum of the variances for two-mode displacement thermal probe state (two-mode thermal coherent state). Fig. (\ref{F5a}) represents SLD, RLD-QCRBs, the max of SLD and RLD-QCRBs,  the HCRB (coincides with the RLD-QCRB), and HD bound as functions of the interaction time $t$, with $N_e=0.5$ is a mean number of photons of the Gaussian environment, and the homodyne angle of beam-splitter has been fixing in $\pi/4$. Fig. (\ref{F5b}) represents the HCRB as a function of the mean number of photon $N_e$ for different values of time $t$. We assume that the overall damping rate $\gamma=1$.}\label{Fig5}
 	\end{figure}
 
 \end{widetext} 
The results plotted in Fig. (\ref{Fig5}) are the obtaining when the two-mode mixed coherent thermal state has applied as a probe state. The left figure (Fig. (\ref{F5a})) represents the SLD, RLD-QCRB, the max of SLD and RLD-QCRBs, the HCRB and the CR bound evaluate via homodyne detection as functions of interaction time $t$. We note that, as showing in Fig. (\ref{F5a}), the HCRB coincides with RLD bound any let the value of time $t$, which means that the parameter model is D-invariant. As well as, as a comparison, in the small values of times $t$, the homodyne detection is an ultimate measurement and the HCRB is tight bound; because we find that the sum of the mean squared error obtained from the homodyne detection is corresponding to the HCRB. Contrariwise, in large values of time $t$, the HCRB renders a better performance than which obtained by homodyne detection measurement. The rights figure (Fig. (\ref{F5b})) shows the behavior of the HCRB as a function of the average number of thermal photons ($\bar{n}=N_e$) for different values of time $t$. One notices that the HCRB is evolving linearly with $\bar{n}$. Thus, one can conclude that the mixed coherent thermal state is not appropriate for estimating the displacement parameters in the presence or the absence of influence environment effects. We can explain that from the expression of the SLD, RLD-QCRBs, HD bound, and the HCRB (Eqs. (\ref{Eq. 47}, \ref{Eq. 48}, \ref{EQQ73}, \ref{Eq. 777})), which they do not depend on the mean energy of the coherent state. It implies that the performance of measurement cannot be controlling from the probe state, and therefore no enhancement can be achieving in the estimation of displacement parameters. These results are similar to those obtained for the pure coherent probe state.

By inspecting the HCRB (TMDV, TMDT) or the upper and bottom bound of HCRB (TMSV, TMST) in Fig. (\ref{Fig6}), we can clarify the results obtained of the various states. Fig. (\ref{F6a}) shows the behavior of the HCRB for TMDV state, the upper and bottom bound of HCRB for the pure TMSV as functions of interaction time $t$, we fixed the squeezing parameter in $0.4$ and the average number of  Gaussian thermal environment photons in $0.5$. We observe that, for TMSV, the upper and bottom bound of HCRB reached a minimum that goes beyond the standard quantum limit (SQL), which is evaluated by applying the vacuum state or a coherent state as a probe state. That is the best result which one can reach in the enhancement of the estimation precision. While, for the displacement vacuum state, the HCRB coincides with the standard quantum limit without exceeding it. That means that the displacement vacuum state did not provide any addition to allowing to go beyond the limit imposed by the classical strategy in estimating the displacement parameters simultaneously. Fig. (\ref{F6b}) represents the behaviors of the  HCRB for the mixed TMDT, the upper and bottom bound of HCRB for the mixed TMST as functions of interaction time $t$. The plotted behaviors show that the upper and bottom bound of HCRB  of the mixed TMST state exceeds the standard quantum limit. It is similar to what obtains in the pure TMSV state. In addition, the behavior of the upper and bottom bound of HCRB  for pure TMSV state remains below the standard quantum limit during a longer time, $t \in \left[ {0;0,7} \right]$, compared with the upper and bottom bound of HCRB  of the mixed TMST state that beat the SQL when $t \in \left[ {0;0,35} \right]$. While, for the mixed TMDT, the behavior of the HCRB takes values above the standard quantum limit, which is worse than that obtained in the pure TMDV state. Therefore, we conclude that the TMDT state and TMDV state are not appropriate to enhance precision in estimating displacement parameters. Can be explained this result by the independence of the probe state on average energy. Contrariwise, the TMST state and TMSV  state are the best archetypes for improving the accuracy of displacement parameter estimation under the effect of a Gaussian thermal environment. That is essentially due to the entanglement encompassed in such states.
 \begin{widetext}
	
	\begin{figure}[H]
		\centering
		\begin{subfigure}{.49\textwidth}
			\centering
			\includegraphics[width=7.35cm]{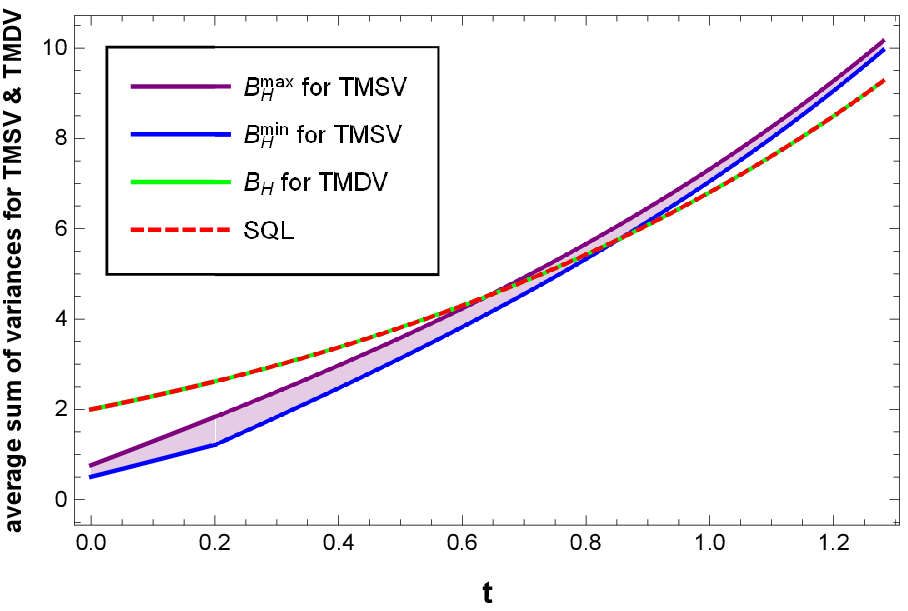}
			\caption{}\label{F6a}
		\end{subfigure}
		\begin{subfigure}{.49\textwidth}
			\centering
			\includegraphics[width=7.35cm]{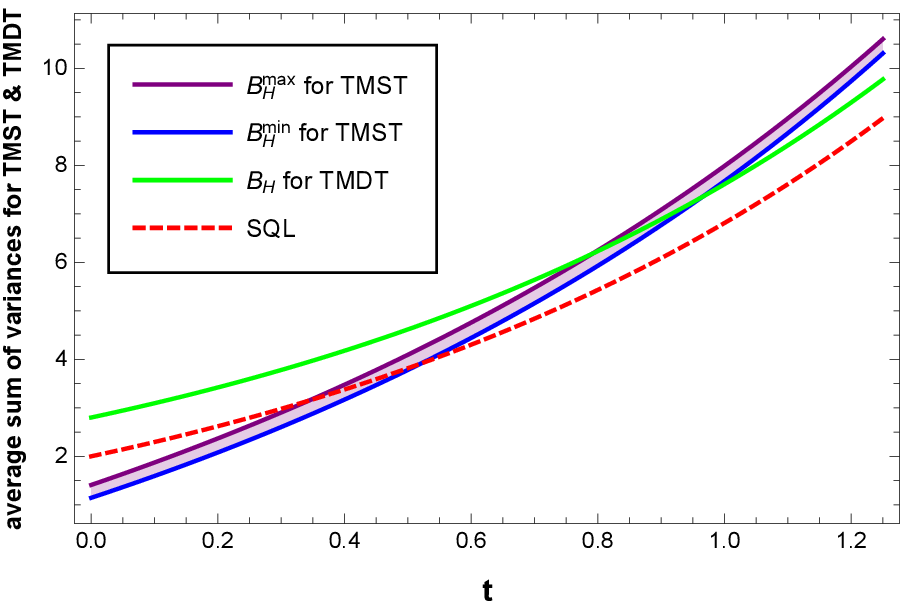}
			\caption{}\label{F6b}
		\end{subfigure}
		\renewcommand\thefigure{\arabic{figure}}
		\addtocounter{figure}{-1}
		\captionsetup{justification=raggedright, singlelinecheck=false, labelfont=sc} \captionof{figure}  {Comparative study of the average sum of variances for the various probe considering states. Fig. (\ref{F6a}) represents the average sum of the variances for pure TMSV and TMDV states as functions of time $t$, with the mean number of photons the Gaussian environment $N_e=0.5$, and te squeezing parameter taking $r=0.4$. The Fig. (\ref{F6b}) represents the average sum of the variances for different mixed TMST and TMDT states as functions of time $t$, with the mean number of photons of Gaussian environment taking $N_e=0.5$, and the squeezing parameter takes $r=0.4$, and the overall damping rate is assumed to equal $\gamma=1$.}\label{Fig6}
	\end{figure}
\end{widetext}
\section{Conclusion and remarks} 
Quantum Gaussian states are one of the building blocks of quantum information with continuous variables systems. Commonly, they are used in various applications due to their ease of generation in the labs also their behavior in the face of losses induced by environmental effects. By using this family of states as a probe state and the homodyne detection measurement, we have investigated, in this paper, a scheme adaptive a protocol to estimate the ultimate precision of the two parameters characterizing the displacement operator modeling the effect of environmental noise. We have studied the limits of possible ultimate measurement by evaluating the upper and bottom bound of HCRB  and the homodyne detection bound. As expected, we noted that the precision of measurement reduced under the effect of the environment. That precisely what one finds when the pure coherent state and mixed coherent thermal state are considered probe states.  Alternatively, if we choose a pure squeezed vacuum state and a  mixed squeezed thermal state as probe states, we get the required enhancement in the estimation precision beyond the standard quantum limit (SQL) even when noises are present. In addition, we have found that the two-mode pure squeezed vacuum states can provide more precision than the two-mode mixed squeezed thermal states with the same squeezing parameter and the same average number of thermal photons. Finally, we emphasize that the obtained results can be explaining by the role of quantum entanglement, which shows again it is a crucial resource to reach the standard quantum limit (SQL), and thus allows to overcome the constraints imposed by lossy evolution in a Gaussian thermal environment.\\
 \vspace*{0.08cm}
\hspace{0.33cm}In view of the obtained results in this work, we believe that, in the future, many perspectives can be considering. The first issue concerns the role of quantum entanglement in improving the accuracy of a metrological protocol in the two-mode Gaussian systems. In this context, we believe that the concept of quantum Fisher information can be employed, as a measure, to quantify the degree of quantum correlations in multi-modes  Gaussian states. In addition, the analysis followed in our paper could be used to analyze the effect of other noisy Gaussian environments. Finally, another prolongation important of this work concerns the multiparameter quantum metrology for states other than Gaussian ones.

\end{document}